\documentclass[structabstract]{aa}

\usepackage{natbib}
 \bibpunct{(}{)}{;}{a}{}{,}

\usepackage{graphicx}

\usepackage{amssymb}
\usepackage{txfonts}
\def\kms{km s$^{-1}$\space}
\def\kmsno{km s$^{-1}$}
\def\micron{$\mu$m\space}

\def\micronno{$\mu$m}
\def\arcsecno{$^{\prime\prime}$}
\def\arcsec{$^{\prime\prime}$\space}

\def\deg{$^{\circ}$\space}

\def\degno{$^{\circ}$}
\def\h2{H$_2$}

\def\cii{[C\,{\sc ii}]\space}
\def\ciino{[C\,{\sc ii}]}
\def\hi{H\,{\sc i}\space}

\def\hii{H\,{\sc ii}\space}
\def\hiino{H\,{\sc ii}}
\def\nii{[N\,{\sc ii}]\space}
\def\niino{[N\,{\sc ii}]}
\def\niii{[N\,{\sc iii}]\space}

\def\oi{[O\,{\sc i}]\space}
\def\oino{[O\,{\sc i}]}

\def\oi{[O\,{\sc i}]\space}
\def\oino{[O\,{\sc i}]}

\def\12co{$^{12}$CO}
\def\13co{$^{13}$CO}
\def\c18o{C$^{18}$O}

\def\C+{C$^+$}
\def\h2{H$_2$}
\def\cm3{cm$^{-3}$}
\def\cm3s{cm$^{-3}$\space}
\def\cm2{cm$^{-2}$}
\def\cm2s{cm$^{-2}$\space}

\begin{document}

\title{The dense warm ionized medium in the inner Galaxy}

\titlerunning{The dense warm ionized medium in the inner Galaxy}

\authorrunning{W. D. Langer et al.}

   \author{W. D. Langer
             \inst{1},
          J. L. Pineda
	           \inst{1},	
          P. F. Goldsmith
	           \inst{1},	        
	            E. T. Chambers
          	 \inst{2},
	            D. Riquelme
          	 \inst{3},
	 L. D. Anderson
	 	\inst{4,}
		\inst{5,}
		\inst{6},
	           M. Luisi
          	 \inst{4,}
		 \inst{5},
	 	M. Justen
		\inst{7},
	            \and 	
          C. Buchbender
          	\inst{7}                 	           
			}


{\institute{Jet Propulsion Laboratory, California Institute of Technology,
              4800 Oak Grove Drive, Pasadena, CA 91109, USA\\
              \email{William.Langer@jpl.caltech.edu}
              \and		               
	 SOFIA-USRA, NASA Ames Research Center, MS 232-12, Moffett Field, CA 94035-0001, USA
	  \and		               	
            Max-Planck-Institut f$\ddot{\rm u}$r Radioastronomie,
		Auf dem H$\ddot{\rm u}$gel 69, 53121 Bonn, Germany
		\and
	Department of Physics and Astronomy, West Virginia University, Morgantown, WV 26506, USA
		\and
		Center for Gravitational Waves and Cosmology, West Virginia University, Chestnut Ridge Research Building, Morgantown, WV 2605, USA
		  \and
		  Green Bank Observatory, P.O. Box 2, Green Bank, WV 24944, USA
		  \and
	  I. Physikalisches Institut der Universit$\ddot{\rm a}$t zu K$\ddot{\rm o}$ln, Z$\ddot{\rm u}$lpicher Strasse 77, 50937
                  K$\ddot{\rm o}$ln, Germany	               		
	         }        	 

   \date{Received December 24, 2020; Accepted April 30, 2021}



\abstract
{Ionized interstellar gas is an important component of the interstellar medium and its lifecycle.  The recent evidence for a widely distributed highly ionized warm interstellar gas with a density intermediate between the warm ionized medium (WIM) and compact \hii regions suggests that there is a major gap in our understanding of the interstellar gas.
}
 {Our goal is to investigate the properties of the dense warm ionized medium  in the Milky Way using spectrally resolved SOFIA GREAT \nii 205 \micron fine-structure lines and Green Bank Telescope hydrogen radio recombination lines (RRL) data, supplemented by spectrally unresolved ${\it Herschel}$ PACS \nii  122\micron data, and spectrally resolved $^{12}$CO.  
 }
 {We observed eight lines of sight (LOS) in the 20\deg $<l <$ 30\deg region in the Galactic plane. We  analyzed spectrally resolved lines of \nii at 205 \micron and RRL observations, along with the spectrally unresolved ${\it Herschel}$ PACS 122 \micron emission, using excitation and radiative transfer models to determine the physical parameters of the dense warm ionized medium.  We derived the kinetic temperature, as well as the thermal and turbulent velocity dispersions from the \nii and RRL linewidths.
}
 {The regions with \nii 205 \micron emission are characterized by electron densities, $n$(e) $\sim$ 10 to 35 cm$^{-3}$, temperatures range from 3400 to 8500 K, and nitrogen column densities $N$(N$^+$) $\sim$ 7$\times$10$^{16}$ to 3$\times$10$^{17}$ cm$^{-2}$.  The ionized hydrogen column densities range from 6$\times$10$^{20}$ to 1.7$\times$10$^{21}$ cm$^{-2}$  and  the fractional nitrogen ion abundance $x$(N$^+$) $\sim$1.1$\times$10$^{-4}$  to 3.0$\times$10$^{-4}$, implying an enhanced nitrogen abundance at a distance  $\sim$ 4.3 kpc from the Galactic Center.   The \nii 205 \micron emission lines coincide with CO emission, although often with an offset in velocity, which suggests that the dense warm ionized gas is located in, or near, star-forming regions, which themselves are associated with molecular gas.
 }
{These dense ionized regions are found to contribute $\gtrsim$ 50\% of the observed \cii intensity along these LOS. The kinetic temperatures we derive are too low to explain the presence of N$^+$  resulting from electron collisional ionization and/or proton charge transfer of atomic nitrogen. Rather, these regions most likely are ionized by extreme ultraviolet (EUV) radiation from nearby star-forming regions or as a result of EUV leakage through a clumpy and porous interstellar medium.  
}


{} 

\maketitle



\section{Introduction}
\label{sec:Section1}

The Galactic interstellar medium (ISM) cycles gas from a diffuse ionized state through succeedingly denser neutral components until dense  molecular clouds form in which young stars are born.  These stars and supernovae generate winds and radiation that disrupt the clouds and ionize the gas, continuing the cycle.  The neutral atomic hydrogen and molecular hydrogen clouds occupy only a small volume of the Galactic disk, $\sim$ 2\%, compared to the ionized gas, which fills much of its volume. A significant fraction of the Galactic disk, $\sim$20\% to 40\%, is filled with low density ionized hydrogen in a warm ionized medium (WIM), also called the diffuse ionized gas (DIG), with an average electron density $\sim$ 0.03 -- 0.08 cm$^{-3}$ and $T_{\rm kin} \sim$8000K \citep{Haffner2009}. The WIM/DIG is estimated to be $\sim$ 90\% of the ionized gas in the Galaxy and $\sim$20\% of the  total gas mass \citep{Reynolds1991}. 

The low density WIM has been studied for nearly six decades since it was proposed as a component of the Galaxy's ISM by \cite{Hoyle1963}.  Pulsar dispersion measurements, faint optical emission lines, and surveys in Hn$\alpha$ emission established that the WIM is widespread throughout the Galaxy \cite[see review by][]{Haffner2009}, in contrast to \hii regions with $n$(e)$\gtrsim$10$^2$ cm$^{-3}$. Bright dense \hii regions are strong sources of emission of the fine structure far-infrared lines of ionized carbon, \ciino, and nitrogen, \niino.  In contrast the WIM can only  be detected in \cii and \nii in absorption towards bright continuum sources with these tracers of ionized gas \citep{Persson2014,Gerin2015}, or in emission along the tangent lines of spiral arms where a long column density of gas has a small velocity dispersion \citep{Velusamy2012,Velusamy2015,Langer2017}.

Thus, the discovery of dense warm ionized gas associated with molecular gas clouds in the inner disk ($l = \pm$ 60\degno) of the Milky Way \citep{Goldsmith2015} was unexpected.  \cite{Goldsmith2015} conducted a survey of  \nii 205 and 122 \micron emission along $\sim$ 150 lines of sight (LOS) in the plane using the {\it Herschel} PACS array and detected relatively strong emission as compared with predictions from the large scale and low angular resolution ($\sim$7\degno) COBE maps, which indicated that \niino, which arises solely from ionized hydrogen gas, was weak \citep{Bennett1994}.   As originally shown by \cite{Rubin1989}, ionized nitrogen requires relatively high electron densities to excite its fine structure levels and would be difficult detect in the WIM with PACS.  The PACS 122 and 205 \micron intensities were used to derive the electron abundances and column densities of the ionized gas.  The densities, $n$(e), were largely in the range 10 to 50 cm$^{-3}$, intermediate between the low density WIM and \hii regions with $n$(e) $\gtrsim$ 10$^2$ cm$^{-3}$, and much lower than ultracompact \hii regions with $n$(e)$>$ 10$^3$ cm$^{-3}$ \citep{Kurtz2005}. Here we investigate the properties of the dense warm ionized gas along eight lines of sight toward the inner Milky Way using spectrally resolved SOFIA GREAT \nii 205 \micron fine-structure lines and Green Bank Telescope hydrogen radio recombination lines (RRL), supplemented by spectrally unresolved {\it Herschel} PACS \nii  205 and 122 \micron data, and CO spectra.  From these data we derive the electron density and temperature, the N$^+$ and H$^+$ column densities, and fractional nitrogen ion abundance.  

In the {\it Herschel} Galactic \nii sparse survey, ten lines of sight observed with PACS were also observed with spectrally resolved 205 \micron emission with  {\it Herschel's} Heterodyne Instrument in the Far-Infrared (HIFI) and found to have $\sim$ 20 distinct velocity components \citep{Langer2016}.   \cite{Pineda2019} analyzed the electron abundance and column density of N$^+$ in these 20 spectral components along with one in the Scutum arm $l \sim$30\deg that had been observed in \nii 205 \micron with SOFIA GREAT \citep{Langer2017}. To derive the electron density, in the absence of spectrally resolved 122 \micron \nii emission, \cite{Pineda2019} developed a technique combining  spectrally resolved Hn$\alpha$ radio recombination lines (RRLs) with the spectrally resolved \nii 205 \micron line.   They found that the  line of sight averaged electron abundances were consistent within a factor of two with the densities derived from the PACS spectrally unresolved 205 and 122 \micron emission, thus validating this approach to determining $n$(e) from the ratio of the \nii to the Hn$\alpha$ RRL intensities.

In addition, \cite{Goldsmith2015} found that this dense warm ionized medium made a significant contribution to the \cii budget, which has important ramifications for understanding the role of \cii in cooling, as a probe of the interstellar medium, and as a tracer of the star formation rate. In contrast to \nii which can only arise from highly ionized gas due to its ionization potential (IP = 14.53 eV) being beyond the Lyman limit, ionized carbon (IP = 11.26 eV) is found in highly ionized and weakly ionized interstellar gas, such as HI clouds, photon dominated regions (PDRs), and low density molecular clouds without CO emission (CO-dark H$_2$ clouds). 

One of the early conclusions from \cii surveys was that, in addition to \cii associated with CO, presumably arising from photon dominated regions, there was widespread \cii not associated with CO \citep{Langer2010,Pineda2013,Langer2014}.  The assumption was that this \cii traces the CO-dark molecular H$_2$ gas and that it represents a significant reservoir of material for eventual star formation in the Milky Way.  \cite{Pineda2013}  found that the average fraction of CO-dark H$_2$, $f_{DG} \sim$ 0.3 in the Milky Way.  This result was in general agreement with earlier estimates of the CO-dark H$_2$ using other indirect measures.   For example, in the Milky Way, Planck found a CO-dark gas fraction $f_{DG} \sim$ 0.5 from excess dust emission \cite[][]{Planck2011} and  \cite{Grenier2005} calculated that $f_{DG} \sim$ 0.3 -- 0.5 using gamma ray emission to trace the hydrogen not detected in CO or HI.  One caveat in these studies of the fraction of the molecular gas traced by CO is that they rely on large scale CO surveys which may not be sensitive enough to detect all the CO emission. However, the analysis of the \cii spectral line surveys had to assume the fraction of its emission that comes from fully ionized regions, due to a lack of observations of spectroscopic probes of the fully ionized hydrogen gas.  

It has been shown recently that emission from ionized nitrogen, N$^+$, in its far-infrared lines, \niino,  can disentangle some of the highly ionized and weakly ionized sources of \cii emission \cite[e.g.,][]{Goldsmith2015,Langer2016,Croxall2017}.  Galactic and extragalactic studies of \nii  indicate that more \cii emission comes from highly ionized gas than predicted by current ISM models, and that the fraction of \cii from highly ionized gas varies among galaxies.  For example, \cite{Abel2006} compares the contributions of PDRs and \hii regions to \cii as a function of the \nii emission (calculated with \hiino --PDR models) available to observations.  Whereas \cii and \nii emission from the \hii regions are well correlated, \cii from the PDRs is not.  Current models of Galactic CO and \cii emission \cite[e.g.,][]{Accurso2017a,Accurso2017b} predict that the majority of galaxies have 60 to 80 percent of their \cii luminosity arising from molecular gas (PDRs), yet analysis of \cii and \nii shows a much wider range within and among galaxies.  Thus, observing \cii alone is insufficient to determine the relative contributions of \hiino, PDRs, and diffuse atomic and diffuse CO--dark H$_2$ clouds.  In addition to \niino, the RRLs of H, He, and C are important probes of conditions in the dense highly ionized \hii regions.  While the electron density can be derived from excitation analysis of the 122 \micron and 205 \micron lines of nitrogen \citep{Oberst2011,Goldsmith2015}, currently spectrally resolved \nii is only available at 205 \micronno. 

The \cii and \nii studies raise important questions about the nature of the dense warm ionized gas, namely: how it forms, what ionizes the nitrogen, what heats this gas, is it transient, and how much does it contribute to the total \cii emission? In this paper our primary goal is to advance our understanding of the physical conditions in the dense warm ionized medium (D-WIM) by combining spectrally resolved \nii 205 \micron and Hn$\alpha$ radio recombination lines, and the PACS observations of spectrally unresolved \nii 122 \micron emission towards eight lines of sight to the inner Galaxy between $l$ = 20\deg to 30\degno, to derive the temperature, electron density, N$^+$ column density, and fractional abundance of N$^+$.  Our secondary goal is to interpret the fraction of  \cii coming from the dense ionized gas by combining its spectrally resolved emission with that of \nii 205 \micronno.   However, the lack of maps in \nii towards the sources limits any discussion of the dynamics of the D-WIM.

This paper is organized as follows.  In Section~\ref{sec:Section2} we describe the observations of \nii and Hn$\alpha$ RRL.    In Section~\ref{sec:Section3} we analyze the properties of the ISM gas traced by \ciino, \niino, and RRL, and in Section~\ref{sec:Section4}  we discuss the characteristics of the D-WIM and the ionization of the gas.  We summarize our results in Section~\ref{sec:Section5}.


\section{Data}
\label{sec:Section2}

We observed eight lines of sight towards the inner Galaxy between $l$ = 21\fdg 9 and 28\fdg 7 at $b$ = 0\deg  in the ionized nitrogen $^3$P$_1 \rightarrow^3$P$_0$  fine structure 205 \micron line, \niino, at high spectral resolution using the 4GREAT instrument on SOFIA. These eight LOS were previously observed in the $^2$P$_{3/2} \rightarrow^2$P$_{1/2}$ 158 \micron fine structure line of ionized carbon, \ciino,  at 1900.537 GHz \citep{Cooksy1986} with {\it Herschel} HIFI  \citep{Langer2010,Pineda2013,Langer2014} and in the \nii 122 and 205 \micron lines with {\it Herschel} PACS at low spectral resolution \citep{Goldsmith2015}.  We also observed hydrogen Radio Recombination Lines along all lines of sight observed in \nii using the Green Bank Telescope as discussed below and described in \cite{Pineda2019}. These lines of sight are marked on Figure~\ref{fig:fig2-1} superimposed on a portion of the Green Bank Telescope (GBT) hydrogen radio recombination line, Hn$\alpha$, Diffuse Ionized Gas Survey (GDIGS) in the Galactic midplane  \citep{Anderson2021}.

 \begin{figure*}[!ht]
 \centering
               \includegraphics[angle=-90,width=18.4cm]{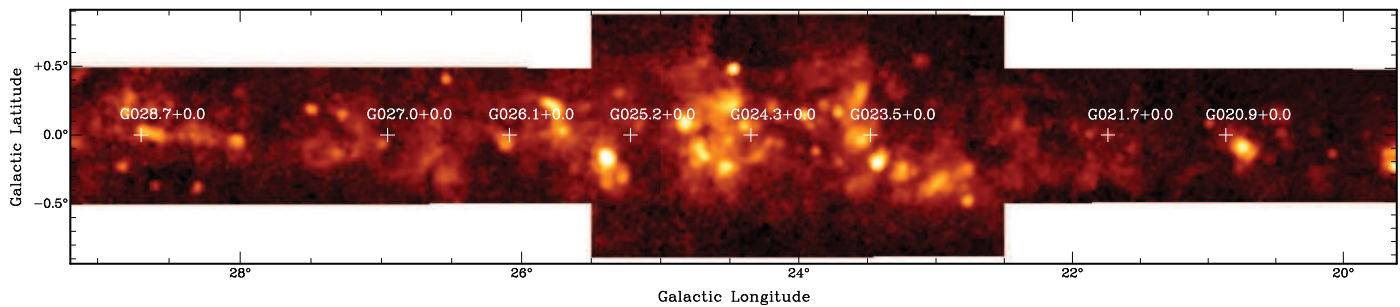}
         \caption{ The location of the eight lines of sight observed in \niino, RRL, and \ciino, indicated $+$ signs, superimposed on a moment 0 integrated intensity map of the Hn$\alpha$ RRL Green Bank Telescope Diffuse Ionized Gas Survey (GDIG; \citep{Anderson2021}). None of the LOS intersect the locations of the brightest compact emission, although many intersect with larger ionized zones. } 
         \label{fig:fig2-1}
              \end{figure*} 

The lines of sight are listed in Table~\ref{tab:Table1} where the first column gives a short label for each line of sight in the form adopted by the GOT C+ survey, GXXX.X+Y.Y, where the first term is the longitude and the second the latitude to one decimal place. The second and third columns give the actual longitude and latitude, $l$ and $b$, as listed in Table 2 of \cite{Goldsmith2015}. The fourth column gives the nominal $V_{LSR}$ used for \nii observations (these are not necessarily the velocities of the peak in emission). The observing details for \nii  are discussed below, while that for the RRL lines are discussed in \cite{Pineda2019}  and the \cii  in \cite{Pineda2013}. The GOT C+ archival data sets are available as a {\it Herschel} User Provider Data Product under KPOT\_wlanger\_1.


\begin{table}																		
\caption{\niino, RRL, and \cii lines of sight.} 
\label{tab:Table1}															
\begin{tabular}{lccc}
\hline
\hline	
LOS ID$^a$ & Longitude$^b$    &  Latitude$^b$  &  V$_{LSR}^c$ \\
 &   &     & ( km s$^{-1}$)\\
   \hline
G020.9+0.0 & 20\fdg 8696 &0\fdg 00 &   50.0    \\
G021.7+0.0 & 21\fdg 7391 &0\fdg 00 &   50.0   \\
G023.5+0.0 & 23\fdg 4783 &0\fdg 00 &   75.0   \\
G024.3+0.0 & 24\fdg 3478 &0\fdg 00 &  75.0   \\
G025.2+0.0 & 25\fdg 2174 &0\fdg 00 &   75.0   \\
G026.1+0.0 & 26\fdg 0870 &0\fdg 00 &  100.0    \\
G027.0+0.0 & 26\fdg 9565 &0\fdg 00 &  100.0   \\
G028.7+0.0 & 28\fdg 6957 &0\fdg 00 &   100.0   \\
 \hline
\end{tabular}
\\
\\
a) We use the GOT C+ naming convention to identify the lines-of-sight ID \citep{Langer2010,Goldsmith2015}.  b) The longitudes and latitudes are from \cite{Goldsmith2015}.  c) The $V_{LSR}$ in the table is that used for the observations and does not necessarily correspond to the peak in emission. 
 \end{table}

\subsection{SOFIA \nii observations}
\label{sec:Subsection2.1}

The \nii line at 1461.1338 GHz $(\lambda$ = 205.178 \micronno) \citep{Brown1994} was observed using GREAT \citep{Heyminck2012,Risacher2016,Duran2020} onboard SOFIA \citep{Young2012} as part of the Guest Observer Cycle 7 campaign  (proposal ID 07\_0009; PI Langer). GREAT was configured with the 4GREAT-HFA \citep{Duran2020} configuration to observe four lines, \niino, \oino, CO(5-4), and CO(8-7). The observations were made on four flights from June 24 to 27, 2019,  from Christchurch, NZ as part of SOFIA's southern deployment. The \nii line was observed with the 4G3 pixel of the 4GREAT receiver, and the corresponding half--power beamwidth for the \nii observations was 19.7\arcsec at 1461.1338 GHz.  The data were processed with the kosma\_ calibrator version 18\_06\_2019 including atmospheric calibration \citep{Guan2012}. The spectra were calibrated to $T_A$ using a forward efficiency of 0.97 and \nii calibrated to main beam temperature with $\eta_{mb}$ = 0.706.  First to third order baselines were removed from the delivered data products. 

Each ON position observation used an OFF position at the same longitude with $b$ = 0\fdg 4, in order to balance baseline quality with the need to have the OFF position as far above the plane as possible.  However, as  \cii and \nii are widespread throughout the Galaxy, and finding a clean OFF position for proper calibration could be a challenge for observations of the inner Galaxy.  Therefore, every OFF position was checked for \nii emission  by observing each LOS at $b$=+0\fdg 4  using an OFF position of 0\fdg 8.  Observations of the Scutum arm in \nii \citep{Langer2017} followed a similar approach but also observed secondary and tertiary OFF positions and found that \nii emission at $b\geq$ 0\fdg 8 is either absent or relatively weak. 

In Figure~\ref{fig:fig2-2} we plot \nii for the lines of sight at $b$ = 0\fdg 0, along with the \cii observations from {\it Herschel} HIFI GOT C+ survey \citep{Pineda2013,Langer2014}.  \nii was detected with high signal-to-noise at five ON positions, with weaker signal-to-noise, at G027.0+0.0,  and marginally detected at two velocities for G021.7+0.0, and was not detected at  G025.2+0.0.  In Appendix~\ref{sec:AppendixA} we plot the \nii spectra for the first OFF positions at $b$ = 0 \fdg 4.

Whenever there was \nii emission at the $b$ = 0\fdg 4 position we corrected the spectra in the ON position by fitting a Gaussian to the spectrum and adding it to the $b$ = 0\fdg 0 spectrum.  Four of the eight LOS had no emission in the OFF position at the level of the rms noise, while three had weak emission, and only G023.5+0.0 had strong emission in the OFF position.  Two examples of the before and after correction spectra are shown in Appendix~\ref{sec:AppendixA}, one for the strongest source, G023.5+0.0, and one for a typical source, G026.1+0.0.  This procedure minimizes the introduction of noise into the target spectra at $b$ = 0\fdg 0.  We then refit the baselines for these $b$ = 0\fdg 0 LOS with a third order polynomial setting a window based on the observed emission.  The typical \nii noise per 2.0 km s$^{-1}$ channel is 0.17 K.

 \begin{figure*}[!ht]
 \centering
               \includegraphics[angle=+0,width=18.5cm]{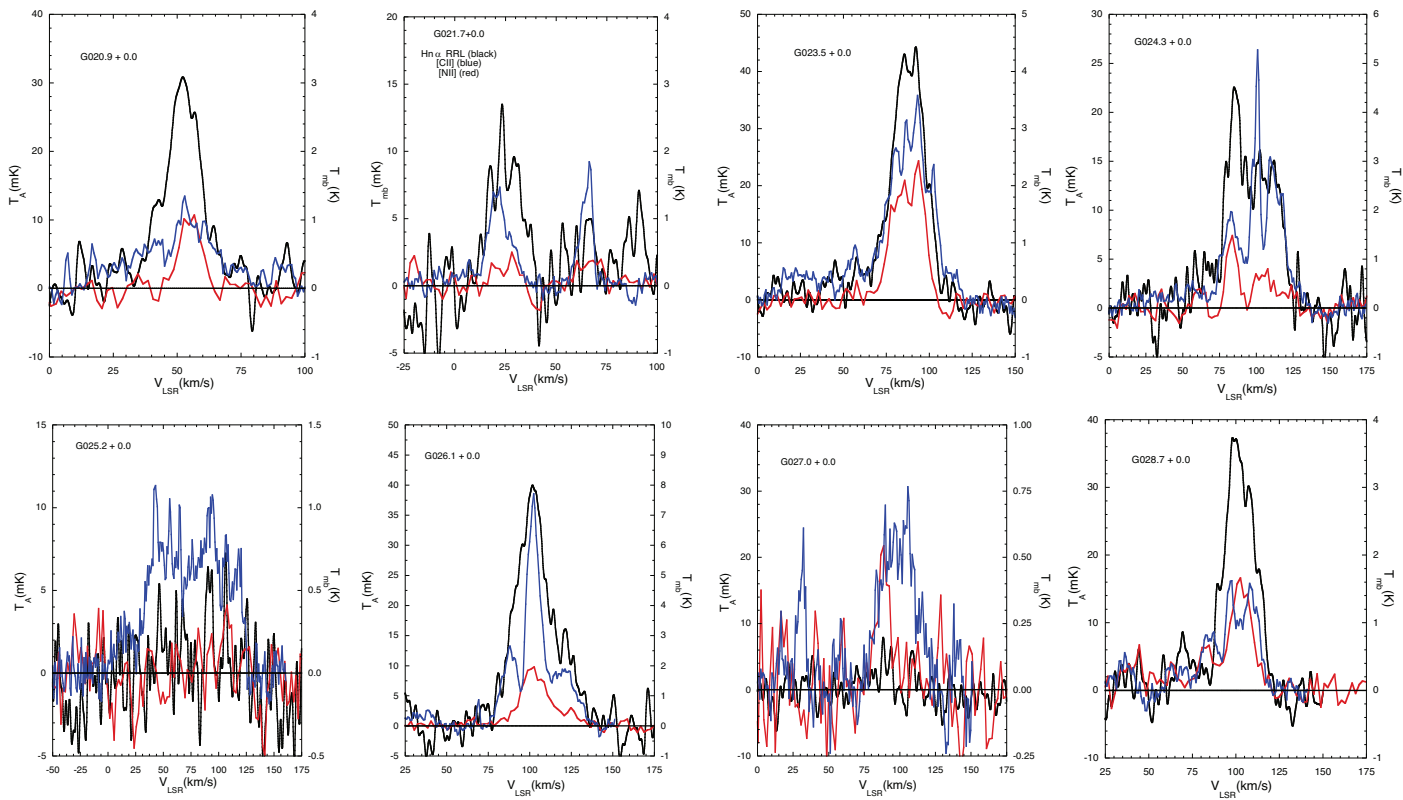}
       \caption{The main beam temperature, $T_{mb}$(K) versus velocity for \nii  (red) and \cii (blue) (right axis and scale) and the corresponding Hn$\alpha$ RRL spectra (black) $T_{mb}$(mK) (left axis and scale) for eight lines of sight (G020.9+0.0 to G028.7+0.0) at $b$=0\fdg 0. \nii was not detected at G025.2+0.0, and was marginally detected towards G021.7+0.0 at two velocities centered on $\sim$ 25 and 67 \kmsno.}  
         \label{fig:fig2-2}
              \end{figure*}

\subsection{Hydrogen recombination lines}
\label{Subsection2.2}

We observed RRLs along all eight LOS in our sample using the Versatile GBT Astronomical Spectrometer (VEGAS) on the Green Bank Telescope (GBT) in X-band (8 -- 10 GHz)  in the position switching observing mode. The angular resolution of the GBT in X-band is 84\arcsec. We used  the same lines and procedures as described in \cite{Pineda2019}, as follows. For each observed direction, we simultaneously measured seven Hn$\alpha$ RRL transitions, H87$\alpha$ to H93$\alpha$, using the techniques discussed in \cite{Bania2010,Anderson2011,Balser2011}, and averaged all spectra together to increase the signal-to-noise ratio using TMBIDL \citep{Bania2016}. All the lines in the band (two polarizations per transition) were averaged together to increase the signal-to-noise ratio and the average frequency is 9.17332 GHz. The GBT data were calibrated using a noise diode fired during data acquisition, and are assumed to be accurate to within 10\% \citep{Anderson2011}. The data were later corrected with a third-order polynomial baseline. We converted the intensities from an antenna temperature to main-beam temperature using a main beam efficiency of 0.94. The typical rms noise of this data is 2.3 mK when smoothed to a 2 km s$^{-1}$ channel. The eight RRL lines are plotted in Figure~\ref{fig:fig2-2} along with \cii and \niino.

 \subsection{GREAT versus PACS 205 \micron data}
\label{sec:Section2-3}

The {\it Herschel Space Observatory} Open Tme Key Program, Galactic Observations of Terahertz C+ (GOT C+)
conducted a sparse survey of velocity resolved \cii along $\sim$500 LOS throughout the disk \citep{Langer2010,Pineda2013,Langer2014} using the HIFI instrument. \cite{Goldsmith2015} did a follow up survey in \nii of 149 LOS observed in \cii by GOT C+. All 149 LOS were observed with {\it Herschel} PACS  in both fine structure transitions of \niino, $^3$P$_2$ -- $^3$P$_1$ at 121.898 $\mu$m, with a resolving power of $\simeq$1000 and $^3$P$_1$ -- $^3$P$_0$ at 205.178 $\mu$m with a resolving power $\simeq$2000. PACS is an array of 25 pixels and its very low spectral resolution provides no velocity information for Galactic LOS, in contrast to the high spectral resolution of HIFI. \nii was detected in 116 LOS at 205 $\mu$m and in 96 of these at 122 $\mu$m, where LOS here means the average of the 25 PACS pixels with an effective angular resolution $\sim$ 47\arcsecno. Details of the PACS data reduction are found in \cite{Goldsmith2015}, where it was noted that the 205 $\mu$m data had calibration uncertainties due to a red filter leak.  With the availability of \nii from SOFIA GREAT it became possible to calibrate the PACS 205 $\mu$m observations. 

To compare the PACS intensities with those of the spectrally resolved \nii data, Pineda (2021, private communication) computed intensities at different angular resolutions by averaging the PACS array pixels weighted with a Gaussian function  with FWHM corresponding to the resolution of SOFIA/GREAT and Herschel/HIFI.  They compared the weighted intensities of the PACS observations with the ten LOS observed with HIFI \citep{Goldsmith2015,Langer2016},  seven LOS observed with GREAT (this paper), and one position in the Scutum arm \citep{Langer2017}.  The half-power beam width of GREAT at 205 \micron is $\simeq$ 20\arcsecno , while the PACS and HIFI beamwidths are $\simeq$ 15\arcsec and 16\arcsecno, respectively.  The PACS weighted pixel average at 205 \micron is lower by $\sim$ 30\%, on average compared with the HIFI and GREAT spectrally resolved 205 \micron intensities. This difference could be due to the different beam sizes, but more likely is a result of the the calibration of the PACS 205 \micron data affected by the red filter leak \cite[c.f.][]{Goldsmith2015}.  

In this paper we have adopted the correction factor suggested by the comparison of PACS with the spectrally resolved data and scaled the PACS 205 \micron intensities by 1.3.  In Table~\ref{tab:Table2} we list the integrated intensities for seven  lines of sight observed with GREAT with those of PACS* where PACS* = 1.3$\times$PACS  from \cite{Goldsmith2015}. Table~\ref{tab:Table2} lists the line of sight, the scaled value of PACS, PACS*, converted to (K km/s) \cite[see][]{Goldsmith2015}, the integrated \nii intensity observed with GREAT in (K \kmsno), and their ratio.  The intensities measured with PACS scaled and GREAT agree reasonably well for the seven positions where \nii is detected with GREAT.  For five of the six LOS $I$(GREAT)/$I$(PACS*) ranges from 0.8 to 1.3, which is what we would expect if the spatial distribution across the PACS pixels does not vary much.  However, the GREAT intensity for LOS G027.0+0.0 is only about 0.6 that observed by PACS.  Along this LOS Pineda (2021, private communication)  found that the ratio in the inner 47\arcsec of the array is much smaller than the average over the array, which suggests that the emission is non-uniform over the PACS array, and probably explains the difference between the GREAT and PACS intensities. G021.7 is one of the weaker sources in \nii detected with PACS.  We have marginal detections, S/N $\sim$ 3 at two velocities, and here use the total integrated intensity which yields a firmer detection, and has about 74\% of the intensity detected with PACS.  In the case of G025.2+0.0 we do not detect any \nii and the 3-$\sigma$ upper limit is only $\sim$ 50\% of the PACS detection.  We have no explanation for the failure to detect \nii along this LOS, but note that the \cii is particularly weak which would be consistent with a weak \nii source. 

 
\begin{table}																		
\caption{Comparing PACS and GREAT 205 \micron integrated intensities.} 
\label{tab:Table2}															
\begin{tabular}{lcccc}	
\hline
\hline
LOS ID$^a$ & $I$(PACS)$^b$    & $I$(PACS*)$^c$  &  $I$(GREAT)& $I$(GREAT)/ \\
 &  (K km/s) &   (K km/s)  &  (K km/s)&$I$(PACS*) \\
   \hline
G020.9+0.0 & 10.5&13.7&   11.0   & 0.81 \\
G021.7+0.0 & 14.1 &18.3 &  10.4$^d$   & 0.74 \\
G023.5+0.0 & 29.4 &38.2 &  49.3   & 1.29 \\
G024.3+0.0 & 20.1 &26.2 &  27.2   & 1.04 \\
G025.2+0.0 & 15.1 &19.6 & $<$ 10.6$^e$   & $<$0.54 \\
G026.1+0.0 & 28.3 &36.7& 43.1 & 1.17\\
G027.0+0.0 & 9.3& 12.1&  7.5   & 0.62 \\
G028.7+0.0 & 19.1 &24.8 &  31.4   & 1.27\\
 \hline
\end{tabular}
\\
\\
a) We use the GOT C+ naming convention to identify the lines-of-sight ID \citep{Langer2010,Goldsmith2015}. b) $I$(PACS) is taken from \cite{Goldsmith2015} c) PACS* is scaled up from the value of PACS reported by \cite{Goldsmith2015} for \nii 205 \micron averaged over all 25 pixels (see text). d) Two features are marginally detected in \nii with GREAT (S/N$\sim$3), however here the total integrated intensity is used which has a S/N$\sim$5.  e) The 3-$\sigma$ upper limit for \niino.\\	
 \end{table}


\section{Properties of the \nii and \cii Gas}
\label{sec:Section3}

In this section we use the \niino, \ciino, and RRL lines to derive the properties of the  ionized gas.  In Figure~\ref{fig:fig2-2} the RRL antenna temperatures are plotted in milli-Kelvin (left axis) while the main beam temperatures  of \cii and \nii are plotted in Kelvin (right axis). The RRL emission was clearly detected in seven of the eight lines of sight with high signal-to-noise, the exception is G025.2+0.0 where there is a marginal detection at $V_{LSR}$ = 92 km s$^{-1}$.     While all eight lines of sight are detected in both \nii transitions using PACS \citep{Goldsmith2015}, only six are clearly detected in the 205 \micron GREAT observations with good S/N.  \nii was marginally detected at G021.7+0.0 and not detected  at G025.2+0.0.   In contrast, every line of sight has detected  \cii emission. 

Here we will use the spectral information combining  RRL, \niino, and \cii emission to derive characteristics of the gas, where possible, such as density,  kinetic temperature,  turbulent velocity dispersion, column density, and fraction of ionized nitrogen.  The analysis of \cii in particular is complicated by foreground or self-absorption of lower excitation gas containing C$^+$.  In principle \nii spectra may also be affected by foreground absorption, and absorption has been detected in \nii against continuum sources \citep{Persson2014}.  However, the abundance of gas phase nitrogen is typically a factor of 2.5 smaller than carbon \citep{Meijerink2005} and absorption of bright \nii emission is likely to be less important. We see no clear evidence of absorption in our \nii data. In contrast to H$^+$ and N$^+$ which are confined to highly ionized gas, C$^+$ is widespread throughout the Milky Way and can be found in many low density, and hence low excitation foreground environments which are capable of absorbing high excitation \cii emission.  For example, \nii and \cii are comparable in intensity for several lines of sight, and \nii is even brighter than \cii in G028.7+0.0 peaking at a local minimum in \ciino.  Similar behavior was seen in other LOS where \cii and \nii were spectrally resolved \citep{Goldsmith2015,Langer2016} and possibly result from foreground absorption.

\subsection{Kinetic temperature and velocity dispersion}
\label{sec:Section3-1}   

The kinetic temperature of the ionized gas is a critical parameter in calculating the properties of the gas as the collisional de-excitation rate coefficients for the fine structure levels of \nii  are temperature dependent \citep{Tayal2011}, as are those of \cii \citep{Tayal2008}.  The temperature is also an important constraint for modeling the heating and cooling of the gas, distinguishing whether collisional ionization is important or not.  Finally, the temperature and density determine the thermal pressure of the gas and factor into whether the gas is thermally confined or will dissipate and on what time scale.

In their derivation of $n$(e) and $N$(N$^+$) \cite{Goldsmith2015} adopted a fixed canonical temperature, 8000K, characteristic of warm ionized gas.   Whereas \cite{Pineda2019} used a fit to the temperature as a function of galactocentric distance derived from the ratio of the hydrogen RRL line-to-continuum emission measurements in \hii regions \cite[e.g.,][]{Balser2015}, which is independent of $n$(e) and proportional to $T_e^{-1.15}$.   \cite{Balser2015} derived electron temperatures for a large sample of \hii regions across the Galaxy and find that they range from $\sim$4500K to $\sim$13,000K with a fit, $T_e$ = 4446 + 467$R_{gal}$(kpc).  The \nii emission studied here are all located in a rather narrow range of Galactic longitude, $\sim$ 20\deg to 30\degno, corresponding to $R_{gal} \sim$ 4 to 6 kpc, which would imply a characteristic temperature $\sim$6300K to 7300K, using the results in \cite{Balser2015}.  However, the temperature gradient derived by \cite{Balser2015}  may not apply to the D-WIM regions studied here, which have much lower electron densities than \hii regions and therefore may have different characteristic temperatures.

There are a number of other techniques to derive the kinetic temperature of the gas.  One is from excitation analysis of two or more levels which have different excitation energies comparable to, or larger than, the kinetic temperature.  However, this approach does not work in the hot highly ionized gas as the two fine structure lines of \nii arise from levels with excitation energies $\sim$ 70 and 188K above the ground state.  These are much lower than the kinetic temperature and so the intensities of these lines are insensitive to temperature in ionized regions with temperatures greater than several hundred K.  Another approach is to compare line widths of species with different mass that emit from the same volume, where the larger the mass contrast the larger the difference in thermal linewidth and thus more readily to determine the thermal temperature. 

Here we take advantage of the factor of 14 difference in mass of N$^+$ and H$^+$ and their different line widths $\Delta V$(RRL) and $\Delta V$\niino, to derive the thermal  and turbulent line widths, and then the thermal temperature.  In using these line widths  to derive the thermal and turbulent velocity dispersion, we neglect the contribution of velocity gradients, opacity, and pressure broadening.  The  \nii opacity at the peak is adapted from \cite{Langer2016},

\begin{equation}
\tau_{10}(N^+)=2.1 \times10^{-17} \frac{f_0}{\Delta V} N_{total}(N^+)\,\,
\label{eqn:eqn3-1}
\end{equation} 

\noindent  where $\Delta V$ is in \kmsno, and $f_0$ is the fractional population in the ground state.  For a column density $N$(N$^+$) = 2$\times$10$^{17}$ cm$^{-2}$ \cite[see Table 2][]{Goldsmith2015}, $T_{th}$ = 8000K, $\Delta V$ = 10 \kmsno, and $f_0 \geq$ 0.5, typical for $n$(e) = 10 to 35 cm$^{-3}$, the opacity $\tau <$ 0.25, and opacity broadening can be neglected (a similar result is shown in Figure 15 in \cite{Goldsmith2015}).   

We cannot rule out line broadening due to systematic motions as we do not have maps in \nii with which to determine the velocity field. We believe that pressure broadening is not likely to be important because there is little evidence of non-Gaussian line shapes for strong lines, in addition a calculation of pressure broadening in a fully ionized hydrogen plasma is beyond the scope of this paper.  A further uncertainty is that the beam widths of the RRL and \nii observations are quite different, which could alter the line shape for non-uniform regions.  In the PACS observations,  at many lines of sight, the \nii emission is roughly constant across the $\sim$ 48\arcsecno $\times$ 48 \arcsec array ($\sim$ 1 pc$^2$ at a distance of $\sim$ 4 kpc), whereas for other LOS it is non-uniform \citep{Goldsmith2015} .  This difference can be seen in Figure 7 of \cite{Goldsmith2015} which compares the HIFI observations to PACS with different pixel averaging.  Here we assume that the different RRL and \nii beamsizes do not significantly affect the line shape, which is less prone to error than the peak intensities. The largest uncertainty, in our view, is the fitting of \nii lines with multiply blended components, as seen in G023.5+0.0 and in several LOS in \cite{Langer2016}.

The solutions are derived in Appendix~\ref{sec:AppendixB} and for N$^+$ and H$^+$ are given by,

\begin{equation}
\Delta V_{th}(H^+) = \bigg (\frac{M_N}{M_N-M_H}\bigg )^{0.5}[\Delta V_{obs}^2(H^+)-\Delta V_{obs}^2(N^+)]^{0.5}\, $\kms$ 
\label{eqn:eqn3-6} 
\end{equation}

\begin{equation}
\Delta V_{turb}(H^+) = (\Delta V_{obs}^2(H^+)-\Delta V_{th}^2(H^+))^{0.5}\, $\kms$
\label{eqn:eqn3-7} 
\end{equation}

\noindent where, $\Delta V_{obs}$ , $\Delta V_{th}$, and $\Delta V_{turb}$ are the observed, thermal, and turbulent Full Width Half Maximum (FWHM) line widths, respectively, and $M_H$ and $M_N$ are the Hydrogen and Nitrogen atomic mass, respectively. In the positions without detection of \nii we can place a strict upper limit on the thermal temperatures from the width of the hydrogen RRL lines.

We fit a Gaussian to the RRL and \nii lines to derive $\Delta V$ which is the Full Width at Half Maximum, FWHM.  In some LOS the RRL and  \nii lines have a single peak and similar line shape  consistent with only one component.  Other LOS spectra show structure that is suggestive of two (or more) components, and we have fit these with two Gaussians.    In the case of \nii the number of components we can fit is limited by the noise level to one or two Gaussians.  In Table~\ref{tab:Table3} we list the Gaussian fits to the RRL and \nii spectra along with the 1-$\sigma$ rms error where $T_p$ is the peak temperature, $V_p$ the velocity of the Gaussian peak, and $\Delta V$, the full width half maximum of the Gaussian.  


\begin{table*}[!htbp]																	
\caption{Spectral line Gaussian parameters.} 
\label{tab:Table3}															
\begin{tabular}{lcccccc}
\hline
\hline	
LOS Label$^a$ & $T_p$(RRL)$^b$    &  $V_p$(RRL)  &  $\Delta V$(RRL)$^c$&$T_p$(\niino)$^d$    &  $V_p$(\niino)$^e$  &  $\Delta V$(\niino)$^c$\\
 &    (mK)   &  (km s$^{-1}$)&(km s$^{-1}$) &    (K)   &  (km s$^{-1}$)&(km s$^{-1}$)\\
   \hline
G020.9+0.0 & 30.3$\pm$0.8 &52.4&   17.2$\pm$0.5   & 1.09$\pm$0.16&55.2&9.6$\pm$1.5 \\
G023.5+0.0-a & 24.0$\pm$0.7&  83.4   & 21.1$\pm$0.6&1.70$\pm$0.17&81.8&13.1$\pm$1.3 \\
G023.5+0.0-b & 25.1$\pm$0.7 & 92.1&   18.2$\pm$0.5   & 1.95$\pm$0.17&92.8&11.7$\pm$1.0\\
G024.3+0.0-a & 20.9$\pm$0.6 &85.1&  14.9$\pm$0.5   & 1.44$\pm$0.16&83.5&8.2$\pm$0.7\\
G024.3+0.0-b & 16.4$\pm$0.6 &106.1&   21.6$\pm$0.8   & 0.72$\pm$0.17& 106.5&19.9$\pm$1.1\\
G026.1+0.0 & 36.8$\pm$0.7 &102.4&   28.0$\pm$0.6   & 1.80$\pm$0.15&102.9&19.9$\pm$1.7\\
G027.0+0.0&4.2$\pm$0.7 &88.5& 30.5$\pm$5.7  &0.48$\pm$0.17&88.4&11.3$\pm$5.0\\
G028.7+0.0 &38.4$\pm$0.7 &102.0&   20.3$\pm$0.4   & 1.49$\pm$0.18&102.7&15.3$\pm$2.1\\
\hline
\end{tabular}
\\
\\
 a) We use the GOT C+ naming convention to identify the lines-of-sight \citep{Langer2010,Goldsmith2015}.  
The actual longitudes and latitudes are given in Table~\ref{tab:Table1}. b) The RRL rms error is for a 1 \kms channel. c) $\Delta V$ is the FWHM of the Gaussian fits. d) The \nii rms error is for a 2 \kms channel. e) The rms errors for $V_p$ are small, less than 0.5 \kms and have no effect on the analysis of the gas properties. 
 \end{table*}

In Table~\ref{tab:Table4} we list the upper bound on the H$^+$ temperature, $T_{upper}$(H$^+$) derived from the RRL Gaussian fits solution to $\Delta V_{obs}$  and Equation~\ref{eqn:eqnA-3}.  We derive, where possible, the thermal and turbulent linewidths using Equations~\ref{eqn:eqn3-6} and \ref{eqn:eqn3-7}, and the thermal temperature of H$^+$.  The H$^+$ thermal temperatures derived from \nii and RRLs lie in the range 3200 to 8500 K.  The two main sources of error for the derived physical parameters are the intensity of the 205 \micron emission (that of the 122 \micron line is much smaller \cite[see Table 2][]{Goldsmith2015} and the temperature, $T_{th}$ because of the dependence of the electron collisional rate coefficients on the population of the \nii levels (see Figure~\ref{fig:fig3-1}). We also list the 1-$\sigma$ rms errors for the solutions.  There is no solution for the one LOS without detectable \nii emission, G025.2+0.0. We also list the thermal and turbulent linewidths, which are typically the same order of magnitude in H$^+$, but for N$^+$ the broadening is dominated by turbulence.

\begin{table*}[!htbp]
\caption{Ionized gas parameters. }
\label{tab:Table4}															
\begin{tabular}{lccccc}
\hline
\hline	
LOS Label$^{a,b}$ & $T_{upper}$(H$^+$)    &  $\Delta V_{th}$(H$^+$) &  $\Delta V_{th}$(N$^+$) & $\Delta V_{turb}$ &$T_{th}$(H$^+$)     \\
 &    10$^3$(K)   & (km s$^{-1}$) &   (km s$^{-1}$)&   (km s$^{-1}$)& 10$^3$(K)  \\
   \hline
G020.9+0.0 & 6.5$\pm$0.4 &14.9$\pm$0.6&3.8$\pm$0.2  & 8.7$\pm$1.9&4.5$\pm$0.4 \\
G023.5+0.0-a &9.8$\pm$0.6 &17.3$\pm$0.4&  4.4$\pm$0.2  &12.2$\pm$1.5& 6.0$\pm$0.3 \\
G023.5+0.0-b &7.2$\pm$0.4 &14.4$\pm$0.3&   3.7 $\pm$0.2 & 11.1$\pm$1.1& 4.2$\pm$0.2 \\
G024.3+0.0-a & 4.9$\pm$0.3 &12.9$\pm$0.2&  3.3$\pm$0.3 &7.4$\pm$0.9&3.4$\pm$0.5 \\
G024.3+0.0-b & 10$\pm$7&13.6$\pm$2.2&  3.5$\pm$0.5  & 16.8$\pm$1.3&3.7$\pm$1.5\\
G026.1+0.0 &  17$\pm$7 &20.5$\pm$0.9&  5.3$\pm$0.3&19.1$\pm$1.9 & 8.5$\pm$0.9\\
G027.0+0.0$^c$ & 24$\pm$7&--&  -- &--&--  \\
G028.7+0.0 &9.0$\pm$0.3 &13.8$\pm$1.8&  3.6$\pm$0.5&14.8$\pm$2.1 &3.9$\pm$0.9\\
\hline
\end{tabular}
\\
\\
a) We use the GOT C+ naming convention to identify the lines-of-sight \citep{Langer2010,Goldsmith2015}. b) The one sigma rms errors are given along with the derived values (see text).   c) The \nii line and RRL are weak and noisy, and the uncertainties on the linewidth too large to determine a reliable value of $T_{th}$(K). \\
 \end{table*}

\subsection{Electron density and N$^+$ column densities}
\label{sec:Section3-2}

In the optically thin limit, which likely applies to \nii \citep{Goldsmith2015}, we can solve exactly for the electron density, $n$(e), and N$^+$ column density, $N$(N$^+$), given the temperature of the ionized gas, and the intensities of the 122 and 205 \micron lines.   \cite{Goldsmith2015} adopted a fixed kinetic temperature of 8000K as typical of the temperature of the ionized gas, whereas we use the kinetic temperatures derived from the line widths in the five LOS observed in \nii 205 \micronno.  

There was not, nor currently is there, a capability to observe a spectrally resolved 122 \micron line.  Therefore, to calculate the electron density from the ratio of the 122 to  205 \micron lines we use the PACS data from \cite{Goldsmith2015}, but the 205 \micron data will be scaled by 1.3 to account for the revised calibration discussed in Section~\ref{sec:Section2-3}.    

The electron density can be obtained from the ratio of the 122  and 205 \micron intensities, and in intensity units of Wm$^{-2}$sr$^{-1}$, we have \cite[Equation 19 in][]{Goldsmith2015},

\begin{equation}
\frac{I_{122}}{I_{205}} = 6.05\frac{f_2}{f_1},\,\,
\label{eqn:eqn3-8}
\end{equation}

\noindent where $f_i$ is the fractional population of level $i$ (as defined in Equation 6 of \cite{Goldsmith2015}) and $I$ is in units of (Wm$^{-2}$sr$^{-1}$).  In Figure~\ref{fig:fig3-1} we plot $I_{122}/I_{205}$ as a function of $n$(e) for temperatures from 2,000K to 20,000K, corresponding to the range of temperatures appropriate to the ionized gas. It can be seen that the values of $n$(e) are sensitive to the kinetic temperature to about  20\% at the low end of the density range, $n$(e) $\sim$ 10 cm$^{-3}$, and about 50\% at the high end, $n$(e) $\sim$ 35 cm$^{-3}$ over the range considered here.  This sensitivity is not due to the excitation energy of the fine-structure levels, which are negligible compared to the kinetic temperature, but to the temperature dependence of the collisional de-excitation rate coefficients \citep{Tayal2011}.  We derived values of $n$(e) for the six LOS with \nii 205 \micron detections with GREAT at high S/N using these temperature sensitive equations  and they are listed in Table~\ref{tab:Table5} along with the distance from the Galactic center, $R_{gal}$ in kpc, calculated following the approach in \cite[][see Section 3.2]{Pineda2019}, and the kinetic temperature. The electron densities we derived using the $T_{\rm kin}$(K) determined in this paper are generally smaller by 50\% than the values derived by \cite{Goldsmith2015} with a fixed $T_{\rm kin}$(K) = 8000K.  The contributions to this reduction are the use of the calibration update described above which increases the PACS 205 \micron intensity, and to generally lower temperatures derived for each region, which change the collisional de-excitation rate coefficients, as seen in Figure~\ref{fig:fig3-1}.  

\begin{figure}
 \centering
               \includegraphics[angle=-0,width=8.5cm]{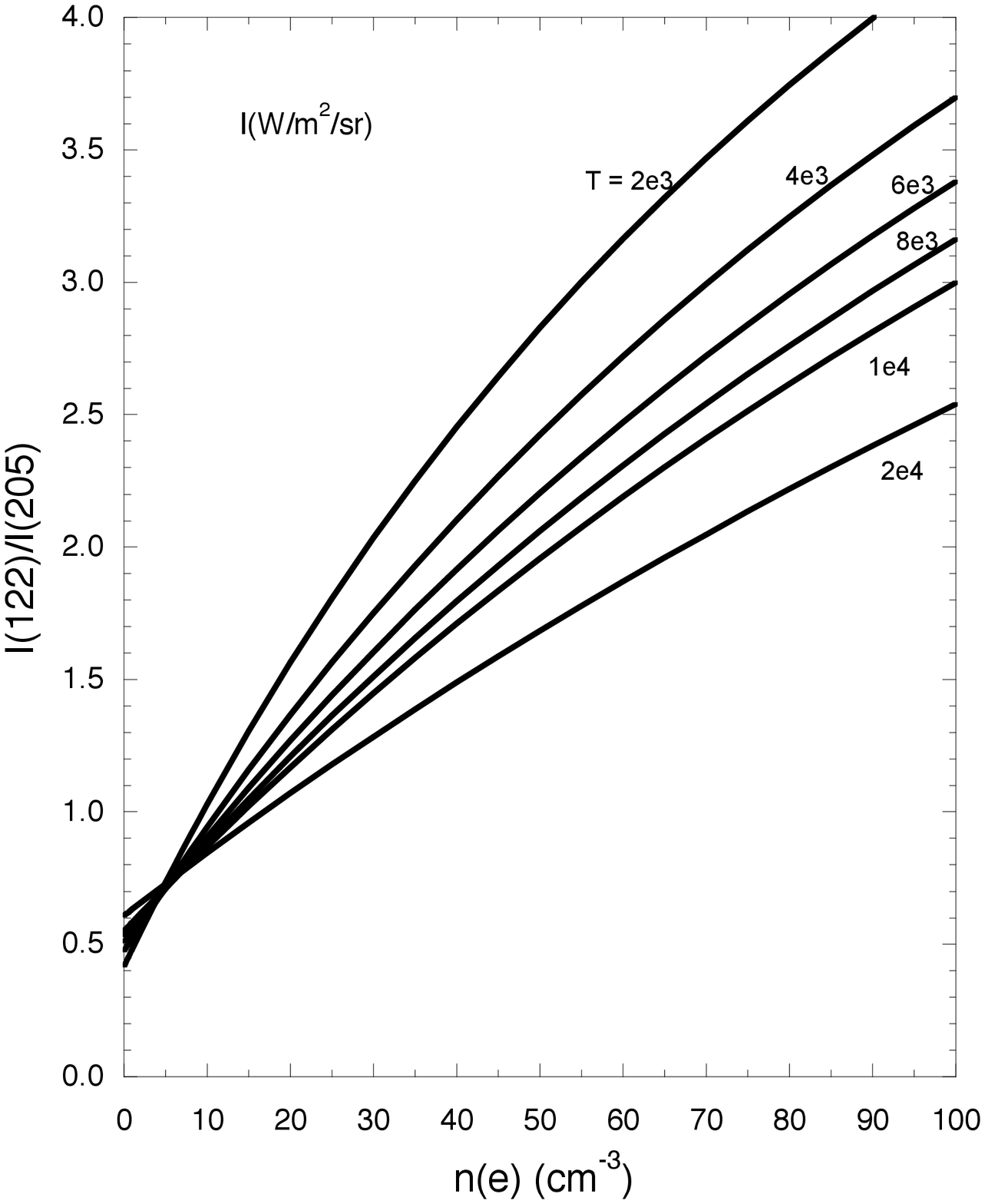}
      \caption{ Intensity ratio of the 122 to 205 \micron lines in W/m$^2$/sr, versus electron density for kinetic temperatures $T_{\rm kin}$(K) ranging from 2,000K to 20,000K.  The curves are labeled by $T_{\rm kin}$(K) and employ the collisional de-excitation rate coefficients of \cite{Tayal2011}. }  
         \label{fig:fig3-1}
         \end{figure}

The column density of N$^+$ can be calculated from the intensity of the 205  \micron line given $T_{\rm kin}$(K) and $n$(e).  We use the GREAT  \nii spectra to calculate the column densities, except for G025.2+0.0 where we do not have a detection and G021.7 where the detection  is marginal. 
In the optically thin limit the column density of $N$($^+$) is given by,

\begin{equation}
N{\rm (N^+)} = 2.0\times 10^{15} \frac{I_{205}{\rm (K\, km/s)}} {f_1}\,\, {\rm (cm^{-2})}\,\,
\label{eqn:eqn3-9}
\end{equation}

\noindent  where $f_1$ is the fractional population of the $^3$P$_1$ level (Equation 6 in \cite{Goldsmith2015}) and the values of $N$(\niino) are given in Table~\ref{tab:Table5} along with those from Table 2 in \cite{Goldsmith2015}.  The ratio of the column densities from both calculations vary considerably with $N_{GREAT}$(\niino)/$N_{PACS}$(\niino) ranging from 0.5 to 2, which reflects the differences in calibration, electron density, and kinetic temperature.   In Table~\ref{tab:Table5} we also list the average size of the \nii emission region defined as,

\begin{equation}
<L> = N({\rm N^+})/n({\rm e}) .\,\,
\label{eqn:eqn3-10}
\end{equation}

\noindent In the following sections we derive the hydrogen ion column density and, assuming N$^+$ and H$^+$ occupy the same volume, the fractional abundance of N$^+$, $x$(N$^+$).


\begin{table*}[!htbp]																		
\caption{Electron densities and N$^+$ column densities.}
\label{tab:Table5}															
\begin{tabular}{lcccccccc}
\hline
\hline	
LOS Label$^a$ &$R_{gal}^b$& $T_{\rm kin}$$^c$ & $I_{205 \mu m }$(\niino)$^d$ &$n$(e)$^e$ &$N$(N$^+$)$^e$  & $<L>$&$n$(e)$^f$ & $N$(N$^+$)$^f$ \\
 & (kpc) & (K) &  (K km/s) &(cm$^{-3}$) &(cm$^{-2}$)  & (pc)&(cm$^{-3}$)& (cm$^{-2}$) \\
   \hline
G020.9+0.0 & 5.0 &4500 &  11.0$\pm$0.9 &16.3$\pm$2.7&(7.3$\pm$1.0)$\times$10$^{16}$ &1.5$\times$10$^{-3}$&30.8& 5.8$\times$10$^{16}$  \\
G023.5+0.0& 4.0& 5100 &49.3$\pm$1.1& 17.4$\pm$1.1&(3.3$\pm$0.2)$\times$10$^{17}$ &6.0$\times$10$^{-3}$& 31.7& 1.6$\times$10$^{17}$  \\
G024.3 +0.0 & 4.3& 3400 &27.2$\pm$1.1&11.2$\pm$1.1& (2.1$\pm$0.2)$\times$10$^{17}$ &6.1$\times$10$^{-3}$& 23.1& 1.3$\times$10$^{17}$  \\
G026.1+0.0 & 3.9& 8500 & 43.1$\pm$1.4& 32.7$\pm$3.9&(2.3$\pm$0.2)$\times$10$^{17}$ &2.1$\times$10$^{-3}$& 48.6& 1.3$\times$10$^{17}$  \\ 
G027.0+0.0&4.3& 8500$^g$& 7.5$\pm$1.1& 14.4$\pm$4.1&(7.5$\pm$1.8)$\times$10$^{16}$ &2.2$\times$10$^{-3}$ & 22.8& 5.9$\times$10$^{16}$  \\
G028.7+0.0 &4.1& 3900 & 31.4$\pm$1.6&20.7$\pm$3.2& (1.8$\pm$0.2)$\times$10$^{17}$ & 2.8$\times$10$^{-3}$& 40.4& 9.6$\times$10$^{16}$ 
\\ 
\hline
\end{tabular}
\\
\\
a) We use the GOT C+ naming convention to identify the lines-of-sight \citep{Langer2010,Goldsmith2015}. b) R$_{gal}$ is calculated using the approach described in \cite[][see Section 3.2]{Pineda2019}. c) If there are temperature solutions to more than one component we use an average value. d) The 1$\sigma$ error for the integral.  e) The errors include those due to uncertainties in $I$(205\micronno) and $T_{th}(K)$.  f) The solutions to $n$(e) and $N$(N$^+$) derived in \cite{Goldsmith2015}.  g) An accurate value of $T_{th}$ could not be derived (see Table~\ref{tab:Table4}) and we adopt the value from G026.1+0.0 for this LOS.\\
 \end{table*}


\begin{table*}[!htbp]																	
\caption{H$^+$ column densities and N$^+$ fractional abundance.}
\label{tab:Table6}														
\begin{tabular}{lccccc}
\hline
\hline	
LOS Label$^a$ &$R_{gal}^b$& $T_{\rm kin}$$^c$ & $I$(RRL)$^d$ & $N$(H$^+$)& $x$(N$^+$) \\
-- & (kpc) & (K) &  (K \kms) &(cm$^{-2}$)   & \\
   \hline
G020.9+0.0 & 5.0 &4500 & 0.488$\pm$0.022 &6.7$\times$10$^{20}$&1.1$\times$10$^{-4}$  \\
G023.5+0.0& 4.0& 5100 &0.963$\pm$0.031& 1.5$\times$10$^{21}$&2.2$\times$10$^{-4}$  \\
G024.3 +0.0 & 4.3& 3400 &0.666$\pm$0.028&8.4$\times$10$^{20}$&2.5$\times$10$^{-4}$  \\
G026.1+0.0 & 3.9& 8500 &  1.029$\pm$0.016& 1.7$\times$10$^{21}$&1.4$\times$10$^{-4}$  \\ 
G027.0+0.0&4.3& 8500$^g$&0.128$\pm$0.018& 7.2$\times$10$^{20}$&1.0$\times$10$^{-4}$ \\
G028.7+0.0 &4.1& 3900 &0.780$\pm$0.17&6.1$\times$10$^{20}$&3.0$\times$10$^{-4}$   \\
\hline
\end{tabular}
\\
\\
a) We use the GOT C+ naming convention to identify the lines-of-sight \citep{Langer2010,Goldsmith2015}. b) R$_{gal}$ is calculated using the approach described in \cite[][see Section 3.2]{Pineda2019}. c) If there are temperature solutions to more than one component we use an average value. d) The 1$\sigma$ error for the integral.\\
 \end{table*}

\subsection{H$^+$ and N$^+$ column densities}
\label{sec:Section3-3}

The H$^+$ column densities, $N$(H$^+$),  were derived from the RRL lines using the approach described in \cite{Pineda2019} and are listed in Table~\ref{tab:Table6}, along with  $N$(N$^+$) from our analysis of the GREAT data.  Also listed in Table~\ref{tab:Table6} are the distance from the Galactic center, $R_{gal}$, electron density, $n$(e), kinetic temperature, $T_{\rm kin}$, the integrated intensity of the \nii observed with GREAT used to derive $N$(N$^+$).   From  $N$(N$^+$) and $N$(H$^+$) we derive the fractional abundance of nitrogen ions, $x$(N$^+$) and these are listed in Table~\ref{tab:Table6}. For reference we also list the solutions for $n$(e) and $N$(N$^+$) derived by \cite{Goldsmith2015} using the PACS pixel averaged intensities of the 122 and 205 \micron lines.

\subsection{Fractional abundance of N$^+$}
\label{sec:Section3-4} 

The fractional abundances of metals as a function of Galactic radius are important parameters for understanding the baryonic lifecycle of the Galaxy. The nitrogen abundance distribution depends on the star formation history of the Galaxy because it is produced by primary and secondary nucleosynthesis in intermediate and massive stars \citep{Johnson2019}.  Optical observations have traditionally been used to trace metallicity in the solar neighborhood, but extinction by grains makes this approach more difficult to probe the inner Galaxy. 
The infrared observations of \ciino,\niino, and \oi are useful probes of the metallicity because they are not readily absorbed by dust and can probe the inner Galaxy where much of star formation takes place.  

Using the derived column densities of N$^+$ and H$^+$ we can evaluate the fractional abundance of nitrogen in the singly ionized state. In Table ~\ref{tab:Table6} we list the fractional abundances $x$(N$^+$) for all lines of sight where we have detected \nii and RRL lines and plot them as a function of $R_{gal}$ in Figure~\ref{fig:fig3-2}.  The fractional abundance of N$^+$ ranges from  1 to 3$\times$10$^{-4}$, which implies an enhanced nitrogen abundance at a Galactic radius about 4.3 kpc, which is an average distance for the \nii emission, and corresponds to the inner edge of the Milky Way's molecular ring.

\begin{figure}
 \centering
               \includegraphics[angle=-0,width=8.5cm]{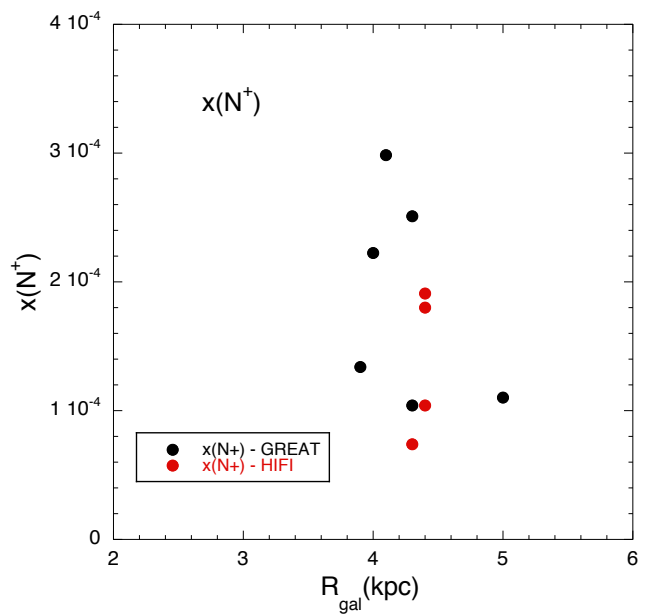}
      \caption{The fractional abundance of N$^+$, $x$(N$^+$), derived from \nii and the RRL emission lines.  The values derived from GREAT are shown in black, while those from HIFI \cite[in Table 3][]{Pineda2019} that are at the same Galactic latitude are plotted in red.  
      }  
         \label{fig:fig3-2}
         \end{figure}

The average value of $x$(N$^+$) $\sim$ 1.9$\times$10$^{-4}$ at $R_{\rm gal}$ $\simeq$ 4.3 kpc is about 2.6 times the nitrogen abundance in the local ISM, $\sim$7.2$\times$10$^{-5}$ \citep{Meijerink2005}, indicating significant enhancement in metallicity in the inner Galaxy.  \cite{Pineda2019} analyzed the nitrogen abundance in the ten lines of sight observed with HIFI in the \cite{Goldsmith2015} survey, using RRLs to derive $N$(H$^+$).   The HIFI data cover a much wider range in $R_{\rm gal}$,  0 to 13 kpc, which made it possible to derive an abundance gradient, 12 + log(N$^+$/H$^+$) = 8.40 - 0.076$R_{\rm gal}$. In the HIFI data \citep[in Table 3][]{Pineda2019} there are four data points at $\sim$ 4.4 kpc, whose average fractional abundance is 1.4$\times$10$^{-4}$,  about 25\% lower than the value derived here.

All the conclusions about the abundance of nitrogen, whether derived from visual or far-infrared observations, assume that nitrogen is entirely in the singly ionized state.  However, as discussed in Section~\ref{sec:Section4}, this depends on the density and temperature of the ionized gas, and the intensity of the extreme UV (EUV) field (photon energies $>$13.6 eV) which regulate the ionization distribution \citep{Langer2015N}.  If nitrogen is in neutral (N) or highly ionized states (mainly N$^{2+}$) then the $x$(N$^+$) values calculated here and in \cite{Pineda2019} are only lower limits.  Indeed there is evidence for doubly ionized nitrogen from observations of  \niii at 57 $\mu$m in diffuse regions of Carina \citep{Mizutani2002}, which could be associated with the D-WIM.  The presence of N$^{2+}$ and higher ionization states would result in an underestimate of the fractional abundance of nitrogen.  However, without  observations of \niii along the lines of sight studied here, we cannot say whether this effect is important here.

\subsection{\cii Emission from highly ionized gas}
\label{sec:Section3-5}

In principle, the \nii emission can be used to determine how much \cii arises from highly and weakly ionized gas because, as discussed above, \nii exists only in highly ionized gas such as the D-WIM, WIM, or \hii regions.  It can be shown that over a wide range of electron densities that the \nii intensity predicts how much \cii should arise from the highly ionized gas as discussed for the Milky Way \cite[e.g.,][]{Goldsmith2015,Langer2016} and in extragalactic sources \cite[e.g.,][]{Croxall2017}.  For effectively optically thin emission the intensity of \cii for ionized gas, $I_{ion}$(\ciino) is given by Equation 4 in \cite{Langer2016},

\begin{equation}
I_{ion}({\rm [CII]})=0.675\frac{f_{3/2}}{f_1}\frac{x({\rm C^+})}{x({\rm N^+})}\  I_{ion}({\rm [NII]})\,\, {\rm (K\, km/s})\,\,
\label{eqn:eqn3-11} 
\end{equation}

\noindent where $f_{3/2}$ and $f_1$ are the fractional populations of the $^2P_{3/2}$ and $^3P_1$ levels of C$^+$ and N$^+$, respectively, and $x$(C$^+$) and $x$(N$^+$) are the fractional abundances of carbon and nitrogen ions. In Figure~\ref{fig:fig3-3} we plot the ratio of of the population of the levels $f_{3/2}$(C$^+$) to $f_1$(N$^+$) as a function of electron density for electron temperatures in the range 2,000 to 10,000K, which are most appropriate for the regions observed in Hn$\alpha$ RRL and \niino.  The ratio uses the collisional rate coefficients for electrons given by  \cite{Tayal2008,Tayal2011}, and the Einstein $A_{ul}$ are from \cite{Schoier2005}.  The ratio $f_{3/2}$(C$^+$) to $f_1$(N$^+$) is remarkably insensitive to electron density and temperature over the range shown in Figure~\ref{fig:fig3-3}.

\begin{figure}
 \centering
         \includegraphics[angle=-0,width=8.5cm]{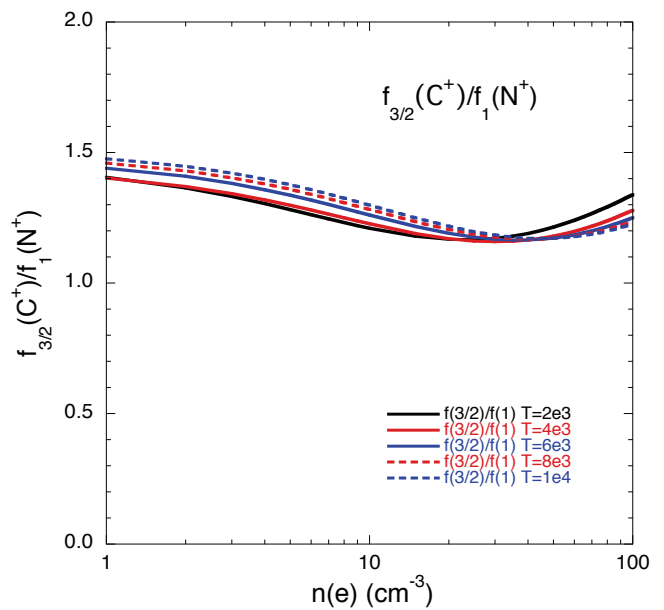}
      \caption{The ratio of the population of the levels $f_{3/2}$(C$^+$) to $f_1$(N$^+$) as a function of electron density for kinetic temperatures in the range 2,000 to 10,000K.}  
         \label{fig:fig3-3}
         \end{figure}

We calculate $I_{ion}$(\ciino) in Equation~\ref{eqn:eqn3-10} using the values of $n$(e) given in Table~\ref{tab:Table7}. Here, for simplicity, we assume that all the carbon and nitrogen are in singly ionized states so $x$(C$^+$)/$x$(N$^+$) is just the elemental carbon to nitrogen ratio (which may not be valid in the presence of a strong EUV flux).  The gas phase abundances of carbon and nitrogen in the inner Galaxy are larger than that in the solar neighborhood \citep{Rolleston2000,Esteban2018}.  In the solar neighborhood, $x$(C)$\approx$1.4$\times$10$^{-4}$ and $x$(N)$\approx$5.2$\times$10$^{-5}$, yields a $x$(C/N) ratio of 2.7.   \cite{Esteban2018} find a nitrogen Galactic radial gradient, log(N/H) = -3.79 -0.059$R_{gal}$, for $R_{gal} >$ 6 kpc, where $R_G$ is in kpc.  Their results are based on optical observations and so cannot directly probe the inner Galaxy.  Extrapolating their relationship to the average radius for \nii emission, $R_G \approx$ 4.3 kpc, yields $x$(N)$\approx$9.0$\times$10$^{-5}$.  For carbon  \cite{Rolleston2000} derived a fractional abundance relationship as a function of Galactic radius, which has been rewritten by \cite{Pineda2013} as $x$(C$^+$) = 5.5$\times$10$^{-4}$10$^{-0.07R_G}$,  and has a value of 2.7$\times$10$^{-4}$ at 4.3 kpc, giving a ratio $x$(C/N) = 3.0, close to the value of 2.9 adopted by \cite{Goldsmith2015}.  Instead of the extrapolation of the optical results to the inner Galaxy we also have the study of \nii and RRL in the inner Galaxy by  \cite{Pineda2019} who derived a slightly different slope from measurements of $R_{gal}$ = 0 to 13.2 kpc.  They found that log(N/H) = -3.6 - 0.076$R_{gal}$, which gives $x$(N) = 1.1$\times$10$^{-4}$ at $R_{gal}$ = 4.5 kpc, about 30\% higher than the \cite{Esteban2018} value.  Combining the N$^+$ results from \cite{Pineda2019} and C$^+$ from \cite{Rolleston2000} yields a carbon to nitrogen ratio, $x$(C/N) =  2.33, about 20\% smaller than the value based on \cite{Esteban2018} and adopted by \cite{Goldsmith2015}.  Here we adopt the lower value for the C$^+$ to N$^+$ ratio using the results of \cite{Pineda2019} combined with \cite{Rolleston2000} which yields a more conservative estimate of the \cii emission predicted to arise from the highly ionized gas using Equation~\ref{eqn:eqn3-11}.  

In Table~\ref{tab:Table7} for each LOS we list the observed line intensities, $I_{tot}$(\ciino) and $I_{tot}$(\niino) in K km/s, the calculated \cii ion intensity, $I_{ion}$(\ciino) using Equation~\ref{eqn:eqn3-11}, and the fraction of \cii emission arising from the ionized and neutral gas. The ratio of $f_{3/2}$/$f_1$ used in Equation~\ref{eqn:eqn3-11} is calculated using the values of $n$(e) and $T_{\rm kin}$(e) derived here (Tables~\ref{tab:Table5} and \ref{tab:Table6}, respectively), although as shown in Figure~\ref{fig:fig3-3} this ratio is not sensitive to either parameter.

For five of the six LOS in which \nii was detected, the intensity of \cii in the highly ionized gas traced by \nii is a significant source of \ciino, representing 50\% to 80\% of the total observed \cii in these LOS.  In one LOS, G028.7+0.0,  the estimated \cii emission from the ionized gas is greater than the observed \cii  intensity.  This result is similar to what has been seen in other LOS in which \nii has been detected  \citep{Goldsmith2015,Langer2016,Croxall2017}.  The fact that the predicted \cii from the highly ionized region is large suggests that a significant fraction of the observed \cii comes from the highly ionized dense ionized warm interstellar medium (D-WIM).  This result is significant because it indicates that the dense warm ionized medium emits \cii comparable to that from PDRs and CO-dark H$_2$ clouds.

In G028.7+0.0 the predicted \cii from the ionized gas is greater than the observed \cii and this deficiency in \cii emission is likely due to foreground  absorption  at $\sim$ 103 km/s where \nii peaks, but \cii shows a clear local minimum.  Foreground absorption of \cii is seen in other LOS \citep{Langer2016,Langer2017} and so appears to be a common feature of the \cii emission.  There is also evidence from [$^{13}$C\,{\sc ii}] detections that \cii has moderate to high opacity in bright \cii sources  \citep{Guevara2020,Kirsanova2020}.  Evidence of deep foreground absorption was established by \cite{Graf2012} from observations of \cii and [$^{13}$C\,{\sc ii}] providing direct evidence of large column densities of cold foreground \cii capable of absorbing background \ciino. However, whether absorption of \cii can explain all of  the emission deficiency or not,  the very large fraction of \cii that appears to arise from the highly ionized gas suggests that a significant fraction of Galactic \cii emission is absorbed by lower excitation foreground gas.

There is one other potential effect on using \nii to calculate the contribution of highly ionized gas to the \cii budget, namely the possibility that the EUV field, or X-rays, produces doubly ionized nitrogen, N$^{2+}$, or even higher ionization states, as seen in the detection of \niii towards highly ionized regions \cite[e.g.,][]{Mizutani2002} and predicted in \hii --PDR models \citep{Abel2005,Kaufman2006}. We discussed the effect of having highly ionized nitrogen on modifying results that measure the fractional abundance of nitrogen in Section~\ref{sec:Section3-3}.  However, the effect of EUV and/or X-rays producing multiply ionized states is less important in calculating the ratio of C$^+$ to N$^+$ because these ionization processes produce  both multiply ionized nitrogen and carbon \citep{Langer2015X}, thus preserving, to first order, the carbon to nitrogen ratio derived from C$^+$ and N$^+$.


\begin{table*}[!htbp]																	
\caption{\cii emission fraction from highly ionized gas.} 
\label{tab:Table7}															
\begin{tabular}{lccccc}
\hline
\hline	
LOS Label$^a$ & $I_{tot}$(\niino)$^b$ & $I_{tot}$(\ciino)$^b$    &  $I_{ion}$(\ciino)    & $I_{ion}$(\ciino)/$I_{tot}$(\ciino) & $I_{neut}$(\ciino)/$I_{tot}$(\ciino)\\
 &    (K km s$^{-1}$)& (K km s$^{-1}$) &  (K km s$^{-1}$)&&\\
  \hline
G020.9+0.0 & 11.0&37.3$\pm$0.8&20.6&0.58&0.42 \\
G023.5+0.0 &49.3&121.3$\pm$0.8&91.9&0.80&0.20 \\
G024.3+0.0 &29.2&107.4$\pm$0.9&56.1&0.53&0.47 \\
G026.1+0.0 &43.1&133.2$\pm$0.7&78.2&0.63&0.37  \\
G027.0+0.0 &7.5&18.1$\pm$0.9&14.5&0.82&0.18  \\
G028.7+0.0 & 31.4&37.1$\pm$0.8&57.7&1.66&$<$0  \\
\hline
\end{tabular}
\\
\\
a) We use the GOT C+ naming convention to identify the lines-of-sight \citep{Langer2010,Goldsmith2015}. b) The rms error is for a 1 \kms channel.\\
\end{table*}


\section{Discussion}
\label{sec:Section4}

The dense warm ionized medium (D-WIM) represents a newly identified state of the ISM  with properties different from those of other highly ionized components of the ISM.  For example, the temperatures we derived for the dense warm ionized gas  range from 3400K to 8500K, comparable to the temperatures derived for dense  \hii regions and the low density WIM.  Whereas the D-WIM electron densities are much larger than those of the WIM and much lower than in \hii regions. We do not yet know the role of the D-WIM in the lifecycle of the ISM, for example, how it forms and evolves. Here we outline some properties of the D-WIM and discuss the likely ionization process sustaining N$^+$.

\subsection{Dense ionized gas component in the ISM}
\label{sec:Section4-0}
 
The \nii emission presented here and in the survey by \cite{Goldsmith2015} primarily trace dense ionized gas rather than the  WIM which has a density too low to produce the observed intensities.  Furthermore, the widespread distribution seen in the PACS survey indicates that the D-WIM is a distinct component of the interstellar medium associated with \cii emission.  All the \nii detections are associated with CO emission, as can be seen in Figure~\ref{fig:fig4-1} which compares the \nii spectra towards the six LOS with the $^{12}$CO(J=1-0) spectra from the GOT C+ survey  \citep{Pineda2013,Langer2014}.  However, not all the CO detections have \nii emission characteristic of the D-WIM.  In addition,  in most cases where \nii and $^{12}$CO are detected there is a velocity shift of \nii with respect to CO, which might indicate a dynamical evolution, although we cannot rule out that some of it could be the result of CO absorption.  

We can also see in the GDIGS maps  (Figure~\ref{fig:fig2-1}) that the \nii LOS do not intercept the brightest emission, probably characteristic of compact \hii sources, but rather are primarily located at the edges of the Hn$\alpha$ emission.  The velocity shift between \nii and CO  is consistent with the D-WIM being associated with star-forming regions which are themselves associated with molecular gas clouds.   This shift indicates that the D-WIM gas traced by \nii are likely  undergoing dynamical evolution, probably a result of energy input from nearby \hii regions seen in the GDIGS survey shown in Figure~\ref{fig:fig2-1}.  However, without maps of \nii we cannot describe the dynamics of the ionized gas.  What is clear is that, unlike the widespread WIM which fills large volumes of the ISM, the D-WIM likely occupies a relatively narrow layer associated with star-forming regions.

The D-WIM could be a density enhancement in the WIM,  as pulsar dispersion measurements suggest that the WIM is clumpy \cite[see review by][]{Haffner2009}.  However,  the \nii spectra presented here are associated in velocity space with \cii and CO, so it seems unlikely that the D-WIM is associated with dense clumps in the WIM.  

Another aspect of the D-WIM is that its thermal pressure $P$ = $n$(e)$T_{\rm kin}$(H$^+$) ranges from $\sim$ 3$\times$10$^4$ to 2$\times$10$^5$ (K cm$^{-3}$), which is much higher than that of the traditional low density WIM, $P \sim$ 800 (K cm$^{-3}$) assuming a WIM uniform electron density,  $n$(e) $\sim$ 0.1 cm$^{-3}$ and temperature, $T_{\rm kin} \sim$ 8000 K.  At these pressures the D-WIM would dissipate rapidly unless it was either formed continuously or gravitationally bound to a molecular cloud.  The expansion would take place on a time scale of the size of the D-WIM layer divided by the sound speed of hydrogen.  From Table~\ref{tab:Table5} the size of the N$^+$ layer, $<L>$(N$^+$) = $N$(N$^+$)/$n$(e),  $\sim$ 10$^{11}$ km and  the sound speed is $\sim$ 5 to 10 \kmsno, yielding a dissipation time  $\sim$ 300 years.  

 \begin{figure*}[!ht]
 \centering
               \includegraphics[angle=-0,width=16.cm]{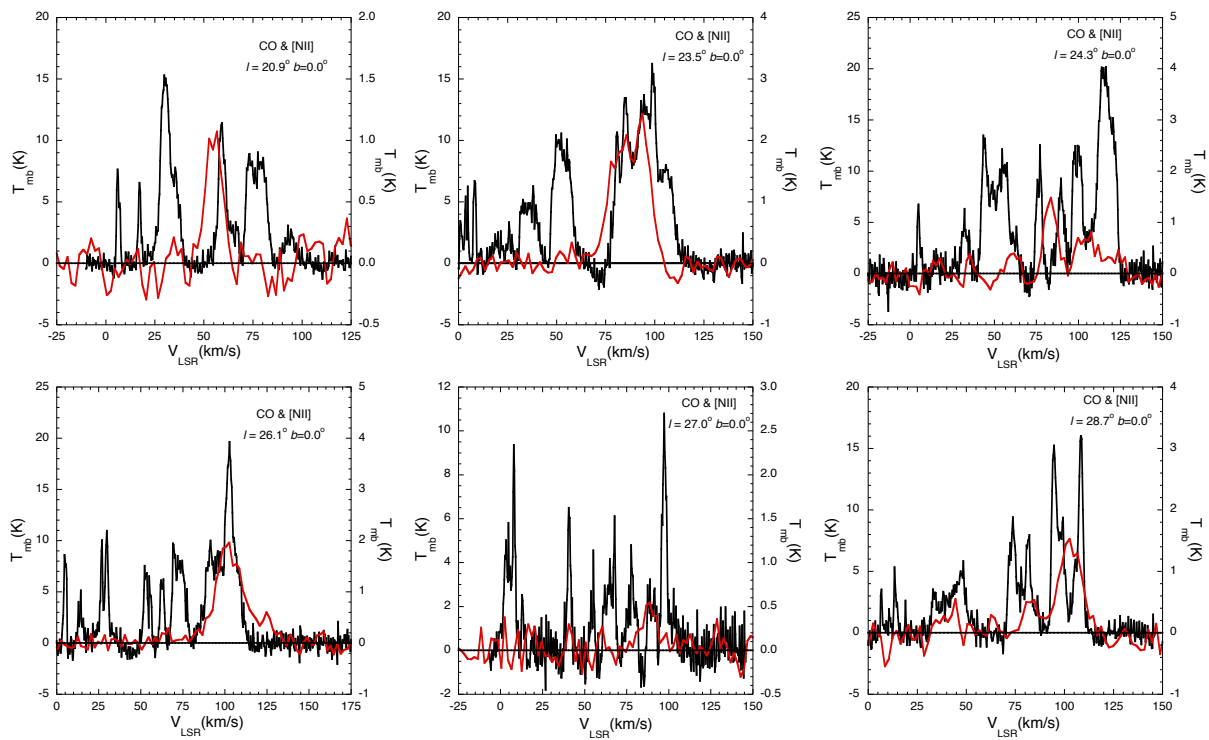}
      \caption{The main beam temperature, $T_{mb}$(K), versus velocity for \nii spectra (red) and $^{12}$CO(1-0) (black) for the six LOS where \nii was clearly detected. }  
         \label{fig:fig4-1}
              \end{figure*} 

\subsection{Ionization of nitrogen}
\label{sec:Section4-1}

The presence of dense ionized nitrogen bears on the ionizing environment as only three processes are important in converting N$\rightarrow$N$^+$; these are  electron collisional ionization, proton charge transfer, and extreme ultraviolet (EUV) photoionization.  In the absence of EUV the first two processes are temperature sensitive but the fractional abundance of N$^+$ is essentially independent of electron density as electron recombination is the dominant loss mechanism in a fully ionized gas (unless the temperature is so high that N$^+$ can be collisionally ionized to N$^{2+}$). If EUV photoionization dominates the production of N$^+$ then the fractional abundance is sensitive to the electron density through the recombination mechanism.  Thus, knowledge of the density and temperature of the dense ionized gas is essential to understand the formation and destruction of nitrogen ions.  Furthermore, the hardness of the EUV, its intensity as a function of wavelength will determine whether higher ionization states exist due to photoionization of N$^+ \rightarrow$ N$^{2+}$. 

\cite{Geyer2018} addressed the ionization of the  D-WIM based solely on electron collisional ionization of N and found that they required a kinetic temperature of $\sim$ 19,000K to explain the presence of dense ionized gas as given by the PACS results \citep{Goldsmith2015}.  A high kinetic temperature is necessary for collisional ionization due to the large ionization potential for e + N $\rightarrow$ N$^+$ + 2e of 14.534 eV.  The kinetic temperatures we derive are clearly in conflict with their conclusions.  However, we note that they neglected to include proton charge exchange and EUV photoionization of nitrogen as possible production of N$^+$.  We have previously discussed electron collisional and proton charge exchange ionization of nitrogen and its sensitivity to the temperature of the gas \citep{Langer2015N}.  Below we consider each of these processes separately to understand the likely source of nitrogen ionization in the D-WIM.
 
\subsubsection{Collisional ionization of nitrogen}
\label{sec:Section4-1-1}

If we consider that the only processes controlling the ionization balance of nitrogen are electron collisional ionization, e + N $\rightarrow$ N$^+$ + 2e, and electronic recombination (electronic + di-electronic), N$^+$ + e $\rightarrow$ N + h$\nu$, then we have for the fractional abundance $f$(N$^+$), 

\begin{equation}
f(N^+) = \frac{k_{ion}f(N)}{k_{rec}}, \,
\label{eqn:eqn4-1}
\end{equation}

\noindent independent of the electron abundance and only dependent on the temperature through the reaction rate coefficients for ionization, $k_{ion}$, and electron recombination, $k_{rec}$.  Substituting for the conservation of nitrogen, $f$(N$^+$) + $f$(N) = 1, we have 

\begin{equation}
f(N^+) = \frac{k_{ion}}{k_{ion}+k_{rec}}.
\label{eqn:eqn4-2}
\end{equation}

 The reaction rate coefficients for electron collisional ionization are derived from thermally averaged cross-sections and we use the fits derived by \cite{Voronov1997}.  The recombination rate coefficients for nitrogen are from \cite{Nahar1997}. In Figure~\ref{fig:fig4-2}  we plot the solutions for electron collisional ionization of nitrogen in terms of $f$(N$^+$) as a function of kinetic temperature of the plasma, $T_{\rm kin}$(K), for temperatures up to 3$\times$10$^4$K.  We neglect ionization to higher ion states because the effect is unimportant due to the high ionization potential for N$^+$ $\rightarrow$ N$^{2+}$ of 29.6 eV.  As can be seen in this figure electron collisional ionization is negligible below  10$^4$ K, and it requires a temperature $>$1.6$\times$10$^4$K for nitrogen to be 50\% ionized, and over   2.5$\times$10$^4$ for nitrogen to be essentially fully ionized.

Proton charge exchange ionization of nitrogen, p +N $\rightarrow$ N$^+$ +H,  is more important than electron collisional ionization at low temperatures because the ionization barrier is the difference in ionization potential between H$^+$ (13.598 eV) and N (14.534 eV), $\Delta$IP = 0.936 eV, or in units of Kelvins, 1.09$\times$10$^4$K.  If the only important collisional processes are proton charge exchange and electron collisional ionization, then we can solve for $f$(N$^+$) in a similar to  Equation~\ref{eqn:eqn4-2}, but replacing $k_{ion}$ with  the charge exchange collisional reaction rate coefficient, $k_{cx}$,

\begin{equation}
f(N^+) = \frac{k_{cx}}{k_{cx}+k_{rec}}\,\,
\label{eqn:eqn4-3}
\end{equation}

\noindent where we use the fits to $k_{cx}$(N$^+ H \rightarrow$ N + H$^+$) as a function of $T_{\rm kin}$ given by \cite{Kingdon1996} multiplied by exp(-$\Delta$IP/$T_{\rm kin}$) to solve for the H$^+$ + N rate coefficients.  In Figure~\ref{fig:fig4-2} we plot the fractional abundance, $f$(N$^+$), when only proton exchange and electron recombination are important (red line).   It can be seen that proton charge exchange is much more effective than electron collisional ionization in producing N$^+$ and begins to ionize the nitrogen measurably above 5000K.  By 10$^4$K the nitrogen is about 40\% ionized, but reaches a maximum of about 50\% at 16,000K. We also plot $f$(N$^+$) when both electrons and proton reactions are considered (black curve), showing how the combined processes keep nitrogen partially ionized above 10$^4$K, but still require temperatures of order 20,000K to reach a nearly fully ionized state.  In order to sustain nearly fully ionized nitrogen in the H$^+$ plasma at the temperatures we measured in \niino, $T_{\rm kin} \leq$ 10$^4$ K, requires EUV photoionization.

\begin{figure}
 \centering
         \includegraphics[angle=-0,width=8.0cm]{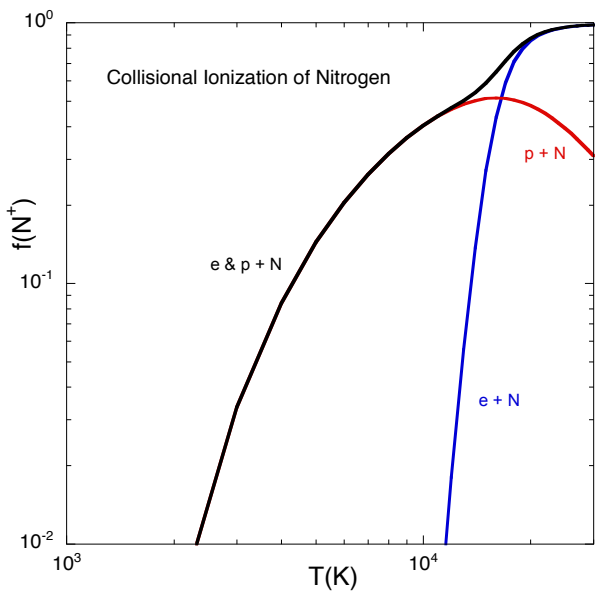}
      \caption{The fractional abundance of ionized nitrogen, $f$(N$^+$), when the only operative processes are electron collisional ionization and recombination (blue) as a function of the plasma temperature, $T_{\rm kin}$(K), compared to $f$(N$^+$) when the only collisional processes are proton charge exchange and recombination (red).  In a realistic case both processes occur simultaneously and the corresponding behavior for $f$(N$^+$) is plotted as the black curve.  }  
         \label{fig:fig4-2}
         \end{figure}

\subsubsection{Photoionization of nitrogen}
\label{sec:Section4-1-2}

Here we explore how EUV photons can keep nitrogen ionized in the D-WIM regions with moderate electron densities.  Young hot stars are sources of ultraviolet radiation that impact the interstellar environment by ionizing atoms and molecules, dissociating molecules, and heating the dust and gas. The ionization potentials for carbon and nitrogen are 11.2603 and 14.5341 eV, respectively.  Thus carbon is readily photoionized by far-ultraviolet photons ($\lambda >$ 912 $\AA$) while nitrogen photoionization requires extreme ultraviolet photons ($\lambda <$ 912 $\AA$).  The EUV radiation produces highly ionized gas that manifests itself as a surrounding or nearby  \hii region.  Furthermore, the EUV from bright \hii sources can leak through holes or clumpy regions of PDRs and sustain a highly ionized gas \citep{Anderson2015,Luisi2016}.  The photoionization rate of nitrogen  by the EUV field is

\begin{equation}
\zeta_{photo\, ion}(N \rightarrow N^+) = \int_{E_{min}}^{E_{max}} \sigma_i(E) \frac{dJ(E)}{dE} dE\,\, {\rm (s^{-1}})\,\,
\label{eqn:eqn4-4}
\end{equation}

\noindent where $dJ(E)/dE$ is the EUV spectral energy distribution in units of photons cm$^{-2}$ s$^{-1}$ eV$^{-1}$, $\sigma(E)$ is the photoionization cross section in cm$^2$, and $E_{mim}$ and $E_{max}$ are the minimum and maximum energies, respectively, in eV.

Whereas the EUV flux from a cluster of massive stars decreases with energy \citep{Abel2005,Kaufman2006}, the cross section for photoionization of nitrogen is relatively flat from threshold to $\sim$25 eV \citep{Samson1990} except for several very narrow resonances.  In addition, the flux of EUV photons decreases with energy and at 20 eV  is roughly a third of that at the Lyman limit at 13.6 eV \citep{Abel2005,Kaufman2006,Sternberg2003}, decreasing to about 10\% at 25 eV. Thus, if we restrict the integral to $E_{min}$ = 14.534 to 20 eV,  to a good approximation,  we can take $\sigma$(E) to be constant $\sigma_0$  $\simeq$ 1.1$\times$10$^{-17}$ cm$^2$, and 

  \begin{equation}
\zeta_{photo\, ion}(N \rightarrow N^+) \simeq \sigma_0 \int_{E_{min}}^{E_{max}} \frac{dJ(E)}{dE} dE\,\, {\rm (s^{-1}})\,\,
\label{eqn:eqn4-5}
\end{equation}

\noindent where term in the integral is just the EUV photon flux between $E_{min}$ and $E_{max}$, 

\begin{equation}
\phi_{EUV} =  \int_{E_{min}}^{E_{max}} \frac{dJ(E)}{dE} dE\,\, {\rm (s^{-1}}).\,\,
\label{eqn:eqn4-6}
\end{equation}

\noindent Thus the nitrogen ionization rate is  given by,

\begin{equation}
\zeta_{photo\, ion}(N \rightarrow N^+) \simeq 1.1\times 10^{-17} \phi_{EUV}\,\, {\rm (s^{-1}}).\,\,
\label{eqn:eqn4-7}
\end{equation}

\noindent More sophisticated approaches can be found in the models of the \hii to PDR transition \citep{Abel2005,Abel2006,Kaufman2006}.   

Using these approximations in the balance equation for photoionization of nitrogen and electron recombination of N$^+$, and neglecting other processes yields a fractional ionization,

\begin{equation}
f(N^+) = \frac{\zeta_{photo\, ion}}{\zeta_{photo\, ion}+k_{rec}n(e)}
\label{eqn:eqn4-8}
\end{equation}

\noindent and  the solution depends on both electron density, $n$(e), and temperature, $T_{\rm kin}$(K).  To have nearly fully ionized nitrogen we need $\zeta_{photo\, ion} >$ 5$\times$$k_{rec}n$(e), or $\phi_{EUV}$$_{min}$ $\gtrsim$ 4.5$\times$10$^{17}$$k_{rec}$$n$(e). 

In Figure~\ref{fig:fig4-3} we plot $\phi_{EUV}$$_{min}$  (photons/cm$^2$/s) as a function of temperature for $n$(e) = 10 cm$^{-3}$, a representative lower limit on density for the D-WIM, such that the lower bound for the EUV photon flux can easily be scaled to other densities. To have a highly ionized nitrogen plasma requires an EUV photon flux $\gtrsim$  a few$\times$10$^6$ photons/cm$^2$/s at $n$(e)=10 cm$^{-3}$ and scales with density. EUV fluxes of this order are associated with massive star-forming regions \citep{Sternberg2003,Abel2005,Kaufman2006}.

\begin{figure}
 \centering
         \includegraphics[angle=-0,width=7.5cm]{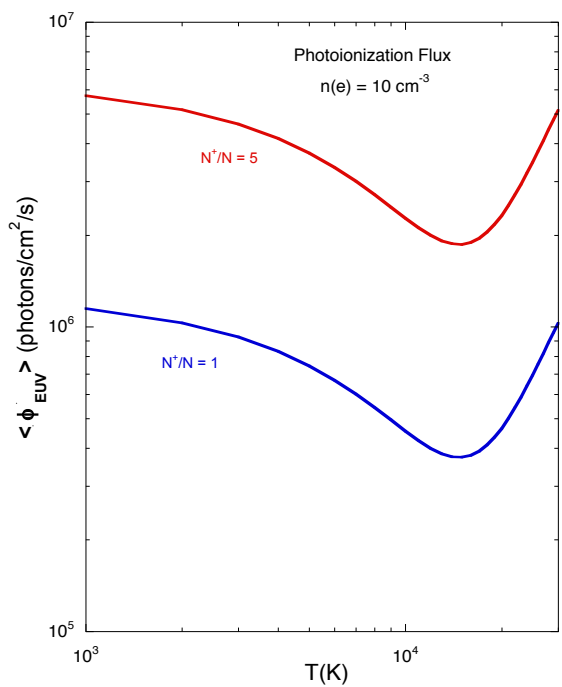}
      \caption{The EUV photon flux required to maintain an ionized fraction of 50\%, N$^+$/N = 1, (blue curve) and 83\%, N$^+$/N =5, (red curve) for a density of $n$(e) = 10 cm$^{-3}$, when the only loss processes is electron recombination as a function of the plasma temperature, $T_{\rm kin}$(K).  The curves are proportional to density. }  
         \label{fig:fig4-3}
         \end{figure}

 \subsection{\cii Emission from the D-WIM}
 \label{sec:Section4-2}
 
  The question of the origin of \cii underlies an important and longstanding quest to understand the composition of the interstellar medium, how \cii emission influences it via cooling the gas, how its emission is affected by star formation heating and ionizing the ISM, and how reliable \cii emission is as a measure of the star formation rate in galaxies.  \hi 21-cm, CO mm and submm rotational lines, and the C$^+$ far-infrared (FIR) fine structure line, \ciino, are among the most important gas tracers that reveal the structure, dynamics, and evolution of galaxies.  Whereas the CO is confined to the shielded regions of molecular clouds and N$^+$ only in highly ionized gas, C$^+$ is found in many of the phases of the ISM in both weakly and highly ionized regions.  The weakly ionized phases are diffuse \hi clouds, diffuse molecular clouds without CO, the so called CO-dark H$_2$ clouds, and photon dominated regions (PDRs) bounding dense molecular clouds.  The highly ionized phases with \cii emission are the warm ionized medium (WIM), with low electron densities, $n$(e) $\lesssim$0.1 cm$^{-3}$, and high temperatures, $T_{\rm kin} \sim$8000 K, dense \hii regions,  the Dense Warm Ionized Medium (D-WIM), and  X-ray dominated regions (XDRs).  Disentangling the \cii contribution from different phases of the ISM is an important goal given the critical importance of \cii in studying Galactic structure, and the star formation rate.
 
 Carbon, with its ionization potential of 11.26 eV, is readily ionized by far-ultraviolet (FUV) radiation ($\lambda <$ 912\AA) that is pervasive throughout the ISM.   Therefore C$^+$ can be found in weakly and highly ionized gas.  In contrast, atomic nitrogen with an ionization potential of 14.5 eV only exists in highly ionized gas that is bathed in extreme ultraviolet (EUV) or X-rays, or is in a hot enough plasma where collisional ionization and proton charge exchange can keep it ionized \citep{Langer2015N,Langer2015X,Langer2016}.  In Section~\ref{sec:Section3-5}  we showed that the \nii emission implies that the D-WIM gas emits a significant fraction of the total \cii along the observed lines of sight.  This result is consistent with prior studies of Galactic and extragalactic \nii and \cii \citep{Goldsmith2015,Langer2016,Croxall2017}.

The conclusions regarding the \cii emission from the D-WIM rest on a number of assumptions, as discussed in Section~\ref{sec:Section3-5}, such as the local gas phase nitrogen to carbon abundance ratio, and the foreground opacity, but less so on temperature and density over the range of conditions derived for the D-WIM.  In fact, as discussed in Section~\ref{sec:Section3-5} there is clear evidence for foreground absorption of \cii  \citep{Langer2016} taken as part of the PACS \nii survey \citep{Goldsmith2015}, and in the SOFIA GREAT Scutum arm observations \citep{Langer2016}.  In contrast there is little or no evidence of foreground absorption of \nii in the HIFI or GREAT data, and using Equation~\ref{eqn:eqn3-1} we estimate that the foreground WIM has an opacity in \nii $\lesssim$ 0.1. However, unlike \niino, \cii is also absorbed by the low density diffuse atomic and molecular clouds. An estimate of the degree of absorption was obtained by \cite{Langer2016} who reconstructed the likely \cii intensity before absorption by fitting the line wings and shoulders of some \cii lines.  He found that the reconstructed  \cii spectra intensity increased by $\sim$ 50\% to 75\%. Thus, as  the \cii LOS observed with SOFIA GREAT do not appear to have large opacities or be significantly absorbed, within a factor of two, and are associated with molecular CO gas \citep{Pineda2013,Langer2014}, it implies that the \nii regions contribute a significant fraction of the observed \ciino.


\section{Summary}
\label{sec:Section5}

The dense warm ionized medium is a recent addition to the known highly ionized interstellar medium components (e.g., WIM, \hiino, hot ionized medium (HIM)). The  N$^+$ fine structure lines at 205   and 122  \micron are important  probes of the D-WIM as, in contrast to \ciino,  they arise only in highly ionized gas due to having an ionization potential above the Lyman limit. The nature and origins of the D-WIM are unclear.  We do not yet understand how it formed, is heated, and can survive against its thermal pressure.  Nor is it clear why it is more prominent in the inner Galaxy (-60\deg $\leq l \leq$ 60\degno), and what is its actual contribution to the \cii Galactic emission.  In this paper we have attempted to shed more light on the D-WIM by characterizing its temperature, electron density, N$^+$ column density, and fractional abundance with respect to the H$^+$ column density.  We have done so by observing spectrally resolved \nii 205 \micron emission with SOFIA GREAT towards eight LOS in the inner Galaxy ($l$ $\sim$ 20\deg to 30\degno) along with hydrogen RRLs observed with the Green Bank Telescope. 

We detected eight \nii components in six of the eight LOS observed and all are associated with \cii emission, however not all \cii components have \nii features. In addition, we had two marginal detections along G027.1+0.0, both aligned with \ciino.  It is important to evaluate the role of \nii in producing \cii throughout the Galaxy for interpreting the  use of \cii and \nii as tracers of the ISM and the baryonic lifecycle.  The PACS survey found a very high correlation between \cii and \nii detections, $>$95\% of 70 LOS, in the inner Galaxy (-30\deg to +30\degno), but a weaker correlation in the outer Galaxy. Overall \nii was detected in 116 LOS out of 149 surveyed.   However, much less is known about the overall association of \nii components with \cii components which requires spectroscopically resolved emission.  Combining the association of \nii components observed here and nine of the ten LOS observed with {\it Herschel} HIFI (excluding the Galactic Center) studied by \cite{Langer2016}, we find that only about 60\% of the \cii components have detectable \nii emission, which is likely due to the lower sensitivity of the single pixel spectroscopic survey compared to the 25 pixel PACS array.

Our analysis of line widths reveals that the D-WIM has temperatures in the range $\sim$ 3400K to 8500K, similar to that of the low density WIM, but not the very high temperatures ($\sim$ 19,000K) suggested by \cite{Geyer2018} from modeling the nitrogen ion fraction of the D-WIM by electron collisional ionization. We find that the inclusion of proton charge exchange increases the ionization fraction, but only photoionization by extreme ultraviolet radiation can explain  a fully ionized dense N$^+$ gas with temperatures consistent with those derived from the comparison of \nii and Hydrogen RRL line widths.  

We recalculated the electron density and N$^+$ column density using the approach in \cite{Goldsmith2015} but using the derived $T_{\rm kin}$(K) for each region rather than adopting a fixed temperature of 8000K or the temperature gradient for the Galaxy derived from dense \hii regions.  We find somewhat lower electron densities, $n$(e) $\sim$ 10 to 30 cm$^{-3}$, about half  those found by \cite{Goldsmith2015}.  The column densities of N$^+$ determined from the \nii 205 \micron line are of order 10$^{17}$ cm$^{-2}$ and they occupy a thin layer about 10$^{16}$ cm thick.  

The fractional abundance of ionized nitrogen, $x$(N$^+$) = $N$(N$^+$)/$N$(H$^+$), was determined for six \nii sources located at $R_{gal}$ $\sim$4.3 kpc and have an average value, $x$(N$^+$) $\sim$1.9$\times$10$^{-4}$ with some degree of scatter.  This value is about 35\% larger than that derived using {\it Herschel} HIFI data, $x$(N$^+$) $\sim$1.4$\times$10$^{-4}$, from four sources at a similar $R_{gal}$ \citep{Pineda2019}. 

The observations of the RRL, \niino, and \cii emission from highly ionized regions indicates that a significant fraction of the \cii emission from the inner Galaxy arises from highly ionized gas, $\sim$ 0.5, and that the emission from weakly ionized gas such as CO-dark H$_2$ clouds and PDRs  contributes less than 50\% of the \cii in contrast to the large fractions previously calculated.  Further, the foreground absorption of \cii seems to be the best explanation for the observed \cii deficit.  To estimate the \cii arising from the dense warm ionized medium we have assumed that all the nitrogen is singly ionized.  However, multiple ionized nitrogen resulting from EUV photoionization only makes this discrepancy worse as the fractional abundance of N$^+$ to C$^+$ now depends on the photoionization of these ions into higher ionization states.  Our analysis reveals that the thermal pressure, $n$(e)$T_{\rm kin}$, in the D-WIM is of order 10$^5$ (K cm$^{-3}$), far exceeding  the pressure of the WIM, $\sim$10$^3$ (K cm$^{-3}$), which presumably is outside the D-WIM.  On the other side of the D-WIM is likely to be a transition to an atomic hydrogen layer and then a PDR. 

Finally, the observations of the D-WIM to date consist mainly of isolated lines of sight that probe the ionized gas in the plane.  To understand its formation  and evolution, its physical and dynamical state, and what it says about this widespread ISM component will require larger scale maps of the Galaxy.



\begin{acknowledgements}
We thank an anonymous referee for a careful reading of the manuscript and numerous suggestions that have improved the interpretation of the observations and the readability of the paper. This research is based in part on observations made with the NASA/DLR Stratospheric Observatory for Infrared Astronomy (SOFIA).  SOFIA is jointly operated by the Universities Space Research Association, Inc. (USRA), under NASA contract NNA17BF53C, and the Deutsches SOFIA Institut (DSI) under DLR contract 50 OK 0901 to the University of Stuttgart.  We thank the SOFIA support teams that made these observations possible.  
The Green Bank Observatory is a facility of the National Science Foundation operated under cooperative agreement by Associated Universities,
Inc. The National Radio Astronomy Observatory is a facility of the National Science Foundation operated
under cooperative agreement by Associated Universities, Inc. {\it Herschel} is an ESA space observatory with
science instruments provided by European-led Principal Investigator consortia and with important participation
from NASA.  LDA and ML acknowledge support from NSF grant AST1812639 to LDA. 
We also thank West Virginia University for its financial support of GBT operations, which enabled some of the observations for this research.
This research was performed in part at the Jet Propulsion Laboratory, California Institute of Technology, under contract with the National Aeronautics and Space Administration. {\copyright}2020 California Institute of Technology. USA Government sponsorship acknowledged. \end{acknowledgements}



\begin{appendix}


\section{\nii at OFF Positions}
\label{sec:AppendixA}

In Figure~\ref{fig:figA-1} we plot \nii for the OFF positions at $b$ = 0\fdg 4.  There are only two \cii OFF positions above the plane at $b$= 0\fdg 5  that have \cii spectra (the other GOT C+ HIFI OFF positions were observed below the plane). These \cii OFF positions are not used here but are shown to indicate that \cii is widespread above the plane, in most cases more so than \niino. Four OFF positions ($b$=0\fdg 4) had no detectable \nii at the level of the three times the rms noise (G020.9+0.4, G021.7+0.4, G024.3+0.4, and G025.2+0.4),  one had strong emission (G023.5+0.4), and the remaining three had weak emission (G026.1+0.4, G027.0+0.4, and G028.7+0.4). 

 \begin{figure*}[!ht]
 \centering
               \includegraphics[angle=-0,width=17.5cm]{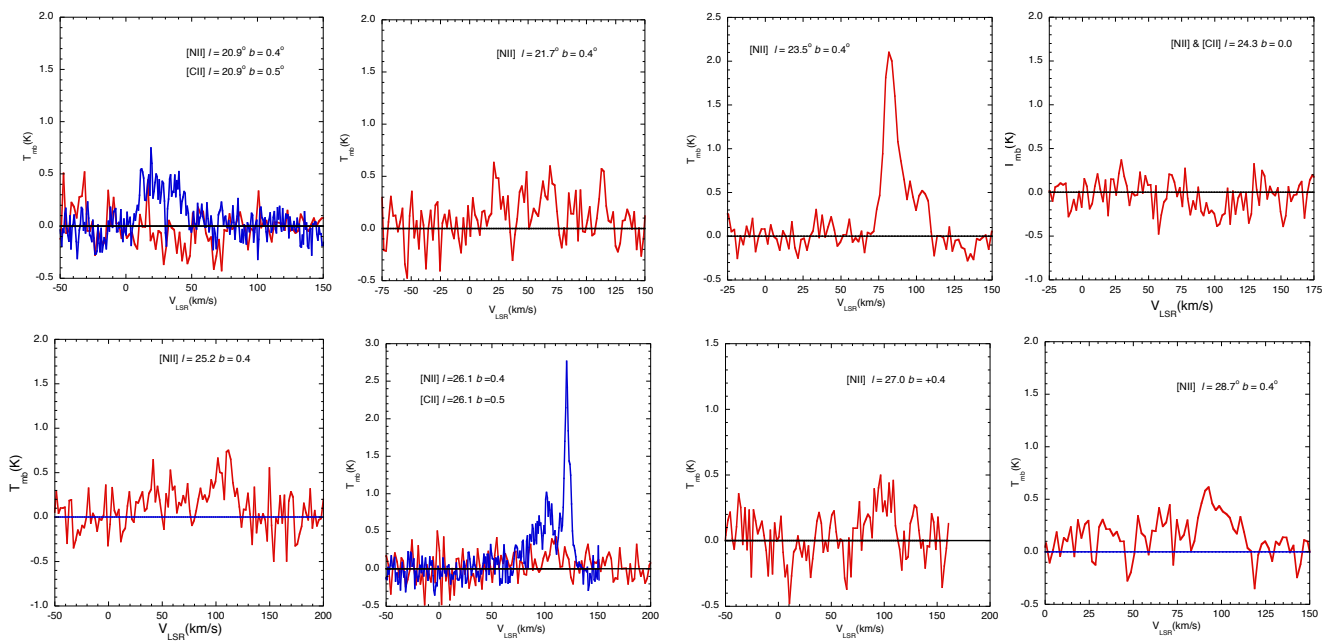}
       \caption{The main beam temperature, $T_{mb}$(K), versus velocity for \nii spectra (red) at the OFF positions for eight lines of sight from 20\fdg 9 to 28\fdg 7 at $b$=0\fdg 4.  Superimposed on the \nii spectra are the corresponding \cii spectra (blue) from {\it Herschel's} GOT C+ survey.  \cii was observed by GOT C+  at only two OFF positions above the plane, $l$=20\fdg 9 and 26\fdg 1, at $b$ = +0\fdg 50.}             
       \label{fig:figA-1}
              \end{figure*}
              
For the four LOS with \nii emission in the OFF position we corrected the $b$=0\fdg 0 spectra by adding back the emission.  In Figure~\ref{fig:figA-2} we show two examples, one for G023.5+0.0, which has the strongest emission at $b$=0\fdg 4, and one for a representative LOS, G026.1+0.0, which has weak emission at the OFF position, $b$=0\fdg 0. The small correction seen for the \nii spectrum at G026.1+0.0 is typical of the other two LOS with weak emission, whereas at G023.5+0.0 the correction is significant. 

 \begin{figure*}[!ht]
 \centering
               \includegraphics[angle=-0,width=18.cm]{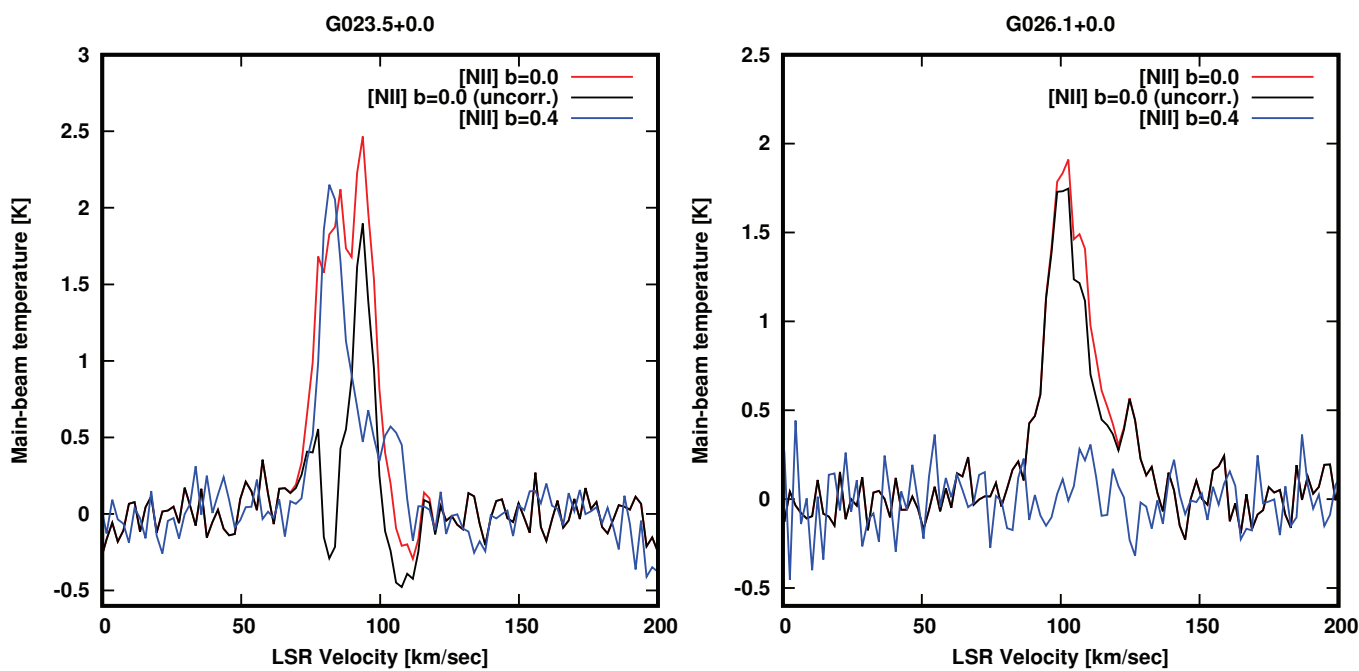}
       \caption{The main beam temperature, $T_{mb}$(K), versus velocity for \nii for two of the four LOS where we had to correct the spectra for emission in the OFF position at $b$ = 0\fdg 4.  The black line is the uncorrected spectrum at $b$ = 0\fdg 0, the blue line is the emission at $b$ = 0\fdg 4, and the red line is the corrected spectrum at $b$ = 0\fdg 0. }               
       \label{fig:figA-2}
                    \end{figure*}

 \section{Thermal and Turbulent Linewidths Derivation}
\label{sec:AppendixB}
 
The velocity dispersion of the RRL and \nii lines can be written as,
\begin{equation}
\Delta V_{o} =( \Delta V_{th}^2 + \Delta V_{turb}^2)^{0.5}\,
\label{eqn:eqnA-1} 
\end{equation}

\noindent where, $\Delta V_{o}$, $\Delta V_{th}$, and $\Delta V_{turb}$ are the observed, thermal, and turbulent Full Width Half Maximum (FWHM) line widths, respectively. For each species, we can write the thermal line width as,

\begin{equation}
\Delta V_{th}(A) = \big (\frac{8\pi kT_{th} ln2}{M_A}\big)^{0.5}\,\, {\rm (cm\, s^{-1})}\,
\label{eqn:eqnA-2} 
\end{equation}

\noindent where A labels H$^+$, C$^+$, or N$^+$, yielding,

\begin{equation}
\Delta V_{th}(H^+) = 0.2141T_{th}^{0.5}\, ($\kmsno$),
\label{eqn:eqnA-3} 
\end{equation}

\begin{equation}
\Delta V_{th}(C^+) = \big (\frac{M_H}{M_C}\big )^{0.5}\Delta V_{th}(H^+) = 0.0618T_{th}^{0.5}\, ($\kmsno$)
\label{eqn:eqnA-4} 
\end{equation}

\begin{equation}
\Delta V_{th}(N^+) = \big (\frac{M_H}{M_N}\big )^{0.5}\Delta V_{th}(H^+) = 0.0572T_{th}^{0.5}\, ($\kmsno$).
\label{eqn:eqnA-5} 
\end{equation}

\noindent  We assume that the \nii (and where we use \ciino) emission arises from the same region as the RRL lines so that we can write Equation~\ref{eqn:eqnA-1} for both the \hii and \nii lines and, assuming that they have the same turbulent velocity dispersion and thermal temperature, $\Delta V_{turb}$(H$^+$) = $\Delta V_{turb}$(N$^+$).  The solution for the thermal and turbulent velocity dispersions for RRL and \nii lines, take the form,

\begin{equation}
\Delta V_{th}(H^+)=\bigg (\frac{M_N}{M_N-M_H}\bigg )^{0.5}[\Delta V_{o}^2(H^+)-\Delta V_{o}^2(N^+)]^{0.5}\, {\rm (km/s)}\,
\label{eqn:eqnA-6} 
\end{equation}

\begin{equation}
\Delta V_{turb}(H^+) = (\Delta V_{o}^2(H^+)-\Delta V_{th}^2(H^+))^{0.5}\, ($\kmsno$).\,\,
\label{eqn:eqnA-7} 
\end{equation}

\noindent In the positions without detection of \nii we can place a strict upper limit on the thermal temperatures from the width of the hydrogen RRL lines.

\end{appendix}



\bibliographystyle{aa}
\bibliography{AA-2020-40223.bib}

\begin{thebibliography}{60}
\expandafter\ifx\csname natexlab\endcsname\relax\def\natexlab#1{#1}\fi

\bibitem[{{Abel}(2006)}]{Abel2006}
{Abel}, N.~P. 2006, \mnras, 368, 1949

\bibitem[{{Abel} {et~al.}(2005){Abel}, {Ferland}, {Shaw}, \& {van
  Hoof}}]{Abel2005}
{Abel}, N.~P., {Ferland}, G.~J., {Shaw}, G., \& {van Hoof}, P.~A.~M. 2005,
  \apjs, 161, 65

\bibitem[{{Accurso} {et~al.}(2017{\natexlab{a}}){Accurso}, {Saintonge},
  {Bisbas}, \& {Viti}}]{Accurso2017a}
{Accurso}, G., {Saintonge}, A., {Bisbas}, T.~G., \& {Viti}, S.
  2017{\natexlab{a}}, \mnras, 464, 3315

\bibitem[{{Accurso} {et~al.}(2017{\natexlab{b}}){Accurso}, {Saintonge},
  {Catinella}, {Cortese}, {Dav{\'e}}, {Dunsheath}, {Genzel}, {Gracia-Carpio},
  {Heckman}, {Jimmy}, {Kramer}, {Li}, {Lutz}, {Schiminovich}, {Schuster},
  {Sternberg}, {Sturm}, {Tacconi}, {Tran}, \& {Wang}}]{Accurso2017b}
{Accurso}, G., {Saintonge}, A., {Catinella}, B., {et~al.} 2017{\natexlab{b}},
  \mnras, 470, 4750

\bibitem[{{Anderson} {et~al.}(2011){Anderson}, {Bania}, {Balser}, \&
  {Rood}}]{Anderson2011}
{Anderson}, L.~D., {Bania}, T.~M., {Balser}, D.~S., \& {Rood}, R.~T. 2011,
  \apjs, 194, 32

\bibitem[{{Anderson} {et~al.}(2015){Anderson}, {Deharveng}, {Zavagno},
  {Tremblin}, {Lowe}, {Cunningham}, {Jones}, {Mullins}, \&
  {Redman}}]{Anderson2015}
{Anderson}, L.~D., {Deharveng}, L., {Zavagno}, A., {et~al.} 2015, \apj, 800,
  101

\bibitem[{{Anderson} {et~al.}(2021){Anderson}, {Luisi}, {Liu}, {Wenger},
  {Balser}, {Bania}, {Haffner}, {Linville}, \& {Mascoop}}]{Anderson2021}
{Anderson}, L.~D., {Luisi}, M., {Liu}, B., {et~al.} 2021, arXiv e-prints,
  arXiv:2103.10466

\bibitem[{{Balser} {et~al.}(2011){Balser}, {Rood}, {Bania}, \&
  {Anderson}}]{Balser2011}
{Balser}, D.~S., {Rood}, R.~T., {Bania}, T.~M., \& {Anderson}, L.~D. 2011,
  \apj, 738, 27

\bibitem[{{Balser} {et~al.}(2015){Balser}, {Wenger}, {Anderson}, \&
  {Bania}}]{Balser2015}
{Balser}, D.~S., {Wenger}, T.~V., {Anderson}, L.~D., \& {Bania}, T.~M. 2015,
  \apj, 806, 199

\bibitem[{{Bania} {et~al.}(2016){Bania}, {Wenger}, {Balser}, \&
  {Anderson}}]{Bania2016}
{Bania}, T., {Wenger}, T., {Balser}, D., \& {Anderson}, L. 2016, {TMBIDL:
  Single dish radio astronomy data reduction package}

\bibitem[{{Bania} {et~al.}(2010){Bania}, {Anderson}, {Balser}, \&
  {Rood}}]{Bania2010}
{Bania}, T.~M., {Anderson}, L.~D., {Balser}, D.~S., \& {Rood}, R.~T. 2010,
  \apjl, 718, L106

\bibitem[{{Bennett} {et~al.}(1994){Bennett}, {Fixsen}, {Hinshaw}, {Mather},
  {Moseley}, {Wright}, {Eplee}, {Gales}, {Hewagama}, {Isaacman}, {Shafer}, \&
  {Turpie}}]{Bennett1994}
{Bennett}, C.~L., {Fixsen}, D.~J., {Hinshaw}, G., {et~al.} 1994, \apj, 434, 587

\bibitem[{{Brown} {et~al.}(1994){Brown}, {Varberg}, {Evenson}, \&
  {Cooksy}}]{Brown1994}
{Brown}, J.~M., {Varberg}, T.~D., {Evenson}, K.~M., \& {Cooksy}, A.~L. 1994,
  \apjl, 428, L37

\bibitem[{{Cooksy} {et~al.}(1986){Cooksy}, {Blake}, \& {Saykally}}]{Cooksy1986}
{Cooksy}, A.~L., {Blake}, G.~A., \& {Saykally}, R.~J. 1986, \apjl, 305, L89

\bibitem[{{Croxall} {et~al.}(2017){Croxall}, {Smith}, {Pellegrini}, {Groves},
  {Bolatto}, {Herrera-Camus}, {Sand strom}, {Draine}, {Wolfire}, {Armus},
  {Boquien}, {Brandl}, {Dale}, {Galametz}, {Hunt}, {Kennicutt}, {Kreckel},
  {Rigopoulou}, {van der Werf}, \& {Wilson}}]{Croxall2017}
{Croxall}, K.~V., {Smith}, J.~D., {Pellegrini}, E., {et~al.} 2017, \apj, 845,
  96

\bibitem[{{Dur{\'a}n} {et~al.}(2020){Dur{\'a}n}, {G{\"u}sten}, {Risacher},
  {G{\"o}rlitz}, {Klein}, {Reyes}, {Ricken}, {Wunsch}, {Graf}, {Jacobs},
  {Honingh}, {Stutzki}, {de Lange}, {Delorme}, {Krieg}, \& {Lis}}]{Duran2020}
{Dur{\'a}n}, C.~A., {G{\"u}sten}, R., {Risacher}, C., {et~al.} 2020, arXiv
  e-prints, arXiv:2012.05106

\bibitem[{{Esteban} \& {Garc{\'\i}a-Rojas}(2018)}]{Esteban2018}
{Esteban}, C. \& {Garc{\'\i}a-Rojas}, J. 2018, \mnras, 478, 2315

\bibitem[{{Gerin} {et~al.}(2015){Gerin}, {Ruaud}, {Goicoechea}, {Gusdorf},
  {Godard}, {de Luca}, {Falgarone}, {Goldsmith}, {Lis}, {Menten}, {Neufeld},
  {Phillips}, \& {Liszt}}]{Gerin2015}
{Gerin}, M., {Ruaud}, M., {Goicoechea}, J.~R., {et~al.} 2015, \aap, 573, A30

\bibitem[{{Geyer} \& {Walker}(2018)}]{Geyer2018}
{Geyer}, M. \& {Walker}, M.~A. 2018, \mnras, 481, 1609

\bibitem[{{Goldsmith} {et~al.}(2015){Goldsmith}, {Y{\i}ld{\i}z}, {Langer}, \&
  {Pineda}}]{Goldsmith2015}
{Goldsmith}, P.~F., {Y{\i}ld{\i}z}, U.~A., {Langer}, W.~D., \& {Pineda}, J.~L.
  2015, \apj, 814, 133

\bibitem[{{Graf} {et~al.}(2012){Graf}, {Simon}, {Stutzki}, {Colgan}, {Guan},
  {G{\"u}sten}, {Hartogh}, {Honingh}, \& {H{\"u}bers}}]{Graf2012}
{Graf}, U.~U., {Simon}, R., {Stutzki}, J., {et~al.} 2012, \aap, 542, L16

\bibitem[{{Grenier} {et~al.}(2005){Grenier}, {Casandjian}, \&
  {Terrier}}]{Grenier2005}
{Grenier}, I.~A., {Casandjian}, J.-M., \& {Terrier}, R. 2005, Science, 307,
  1292

\bibitem[{{Guan} {et~al.}(2012){Guan}, {Stutzki}, {Graf}, {G{\"u}sten},
  {Okada}, {Requena-Torres}, {Simon}, \& {Wiesemeyer}}]{Guan2012}
{Guan}, X., {Stutzki}, J., {Graf}, U.~U., {et~al.} 2012, \aap, 542, L4

\bibitem[{{Guevara} {et~al.}(2020){Guevara}, {Stutzki}, {Ossenkopf-Okada},
  {Simon}, {P{\'e}rez-Beaupuits}, {Beuther}, {Bihr}, {Higgins}, {Graf}, \&
  {G{\"u}sten}}]{Guevara2020}
{Guevara}, C., {Stutzki}, J., {Ossenkopf-Okada}, V., {et~al.} 2020, \aap, 636,
  A16

\bibitem[{{Haffner} {et~al.}(2009){Haffner}, {Dettmar}, {Beckman}, {Wood},
  {Slavin}, {Giammanco}, {Madsen}, {Zurita}, \& {Reynolds}}]{Haffner2009}
{Haffner}, L.~M., {Dettmar}, R.-J., {Beckman}, J.~E., {et~al.} 2009, Reviews of
  Modern Physics, 81, 969

\bibitem[{{Heyminck} {et~al.}(2012){Heyminck}, {Graf}, {G{\"u}sten}, {Stutzki},
  {H{\"u}bers}, \& {Hartogh}}]{Heyminck2012}
{Heyminck}, S., {Graf}, U.~U., {G{\"u}sten}, R., {et~al.} 2012, \aap, 542, L1

\bibitem[{{Hoyle} \& {Ellis}(1963)}]{Hoyle1963}
{Hoyle}, F. \& {Ellis}, G.~R.~A. 1963, Australian Journal of Physics, 16, 1

\bibitem[{{Johnson}(2019)}]{Johnson2019}
{Johnson}, J.~A. 2019, Science, 363, 474

\bibitem[{{Kaufman} {et~al.}(2006){Kaufman}, {Wolfire}, \&
  {Hollenbach}}]{Kaufman2006}
{Kaufman}, M.~J., {Wolfire}, M.~G., \& {Hollenbach}, D.~J. 2006, \apj, 644, 283

\bibitem[{{Kingdon} \& {Ferland}(1996)}]{Kingdon1996}
{Kingdon}, J.~B. \& {Ferland}, G.~J. 1996, \apjs, 106, 205

\bibitem[{{Kirsanova} {et~al.}(2020){Kirsanova}, {Ossenkopf-Okada}, {Anderson},
  {Boley}, {Bieging}, {Pavlyuchenkov}, {Luisi}, {Schneider}, {Andersen},
  {Samal}, {Sobolev}, {Buchbender}, {Aladro}, \& {Okada}}]{Kirsanova2020}
{Kirsanova}, M.~S., {Ossenkopf-Okada}, V., {Anderson}, L.~D., {et~al.} 2020,
  \mnras, 497, 2651

\bibitem[{{Kurtz}(2005)}]{Kurtz2005}
{Kurtz}, S. 2005, in IAU Symposium, Vol. 227, Massive Star Birth: A Crossroads
  of Astrophysics, ed. R.~{Cesaroni}, M.~{Felli}, E.~{Churchwell}, \&
  M.~{Walmsley}, 111--119

\bibitem[{{Langer} {et~al.}(2016){Langer}, {Goldsmith}, \&
  {Pineda}}]{Langer2016}
{Langer}, W.~D., {Goldsmith}, P.~F., \& {Pineda}, J.~L. 2016, \aap, 590, A43

\bibitem[{{Langer} {et~al.}(2015){Langer}, {Goldsmith}, {Pineda}, {Velusamy},
  {Requena-Torres}, \& {Wiesemeyer}}]{Langer2015N}
{Langer}, W.~D., {Goldsmith}, P.~F., {Pineda}, J.~L., {et~al.} 2015, \aap, 576,
  A1

\bibitem[{{Langer} \& {Pineda}(2015)}]{Langer2015X}
{Langer}, W.~D. \& {Pineda}, J.~L. 2015, \aap, 580, A5

\bibitem[{{Langer} {et~al.}(2017){Langer}, {Velusamy}, {Goldsmith}, {Pineda},
  {Chambers}, {Sandell}, {Risacher}, \& {Jacobs}}]{Langer2017}
{Langer}, W.~D., {Velusamy}, T., {Goldsmith}, P.~F., {et~al.} 2017, \aap, 607,
  A59

\bibitem[{{Langer} {et~al.}(2010){Langer}, {Velusamy}, {Pineda}, {Goldsmith},
  {Li}, \& {Yorke}}]{Langer2010}
{Langer}, W.~D., {Velusamy}, T., {Pineda}, J.~L., {et~al.} 2010, \aap, 521, L17

\bibitem[{{Langer} {et~al.}(2014){Langer}, {Velusamy}, {Pineda}, {Willacy}, \&
  {Goldsmith}}]{Langer2014}
{Langer}, W.~D., {Velusamy}, T., {Pineda}, J.~L., {Willacy}, K., \&
  {Goldsmith}, P.~F. 2014, \aap, 561, A122

\bibitem[{{Luisi} {et~al.}(2016){Luisi}, {Anderson}, {Balser}, {Bania}, \&
  {Wenger}}]{Luisi2016}
{Luisi}, M., {Anderson}, L.~D., {Balser}, D.~S., {Bania}, T.~M., \& {Wenger},
  T.~V. 2016, \apj, 824, 125

\bibitem[{{Meijerink} \& {Spaans}(2005)}]{Meijerink2005}
{Meijerink}, R. \& {Spaans}, M. 2005, \aap, 436, 397

\bibitem[{{Mizutani} {et~al.}(2002){Mizutani}, {Onaka}, \&
  {Shibai}}]{Mizutani2002}
{Mizutani}, M., {Onaka}, T., \& {Shibai}, H. 2002, \aap, 382, 610

\bibitem[{{Nahar} \& {Pradhan}(1997)}]{Nahar1997}
{Nahar}, S.~N. \& {Pradhan}, A.~K. 1997, \apjs, 111, 339

\bibitem[{{Oberst} {et~al.}(2011){Oberst}, {Parshley}, {Nikola}, {Stacey},
  {L{\"o}hr}, {Lane}, {Stark}, \& {Kamenetzky}}]{Oberst2011}
{Oberst}, T.~E., {Parshley}, S.~C., {Nikola}, T., {et~al.} 2011, \apj, 739, 100

\bibitem[{{Persson} {et~al.}(2014){Persson}, {Gerin}, {Mookerjea}, {Black},
  {Olberg}, {Goicoechea}, {Hassel}, {Falgarone}, {Levrier}, {Menten}, \&
  {Pety}}]{Persson2014}
{Persson}, C.~M., {Gerin}, M., {Mookerjea}, B., {et~al.} 2014, \aap, 568, A37

\bibitem[{{Pineda} {et~al.}(2019){Pineda}, {Horiuchi}, {Anderson}, {Luisi},
  {Langer}, {Goldsmith}, {Kuiper}, {Bryden}, {Soriano}, \&
  {Lazio}}]{Pineda2019}
{Pineda}, J.~L., {Horiuchi}, S., {Anderson}, L.~D., {et~al.} 2019, \apj, 886, 1

\bibitem[{{Pineda} {et~al.}(2013){Pineda}, {Langer}, {Velusamy}, \&
  {Goldsmith}}]{Pineda2013}
{Pineda}, J.~L., {Langer}, W.~D., {Velusamy}, T., \& {Goldsmith}, P.~F. 2013,
  \aap, 554, A103

\bibitem[{{Planck Collaboration} {et~al.}(2011){Planck Collaboration}, {Ade},
  {Aghanim}, {Arnaud}, {Ashdown}, {Aumont}, {Baccigalupi}, {Balbi}, {Banday},
  {Barreiro}, {Bartlett}, {Battaner}, {Benabed}, {Beno{\^\i}t}, {Bernard},
  {Bersanelli}, {Bhatia}, {Bock}, {Bonaldi}, {Bond}, {Borrill}, {Bouchet},
  {Boulanger}, {Bucher}, {Burigana}, {Cabella}, {Cardoso}, {Catalano},
  {Cay{\'o}n}, {Challinor}, {Chamballu}, {Chiang}, {Chiang}, {Christensen},
  {Clements}, {Colombi}, {Couchot}, {Coulais}, {Crill}, {Cuttaia}, {Dame},
  {Danese}, {Davies}, {Davis}, {de Bernardis}, {de Gasperis}, {de Rosa}, {de
  Zotti}, {Delabrouille}, {Delouis}, {D{\'e}sert}, {Dickinson}, {Dobashi},
  {Donzelli}, {Dor{\'e}}, {D{\"o}rl}, {Douspis}, {Dupac}, {Efstathiou},
  {En{\ss}lin}, {Eriksen}, {Falgarone}, {Finelli}, {Forni}, {Fosalba},
  {Frailis}, {Franceschi}, {Fukui}, {Galeotta}, {Ganga}, {Giard}, {Giardino},
  {Giraud-H{\'e}raud}, {Gonz{\'a}lez-Nuevo}, {G{\'o}rski}, {Gratton},
  {Gregorio}, {Grenier}, {Gruppuso}, {Hansen}, {Harrison}, {Helou},
  {Henrot-Versill{\'e}}, {Herranz}, {Hildebrandt}, {Hivon}, {Hobson}, {Holmes},
  {Hovest}, {Hoyland}, {Huffenberger}, {Jaffe}, {Jones}, {Juvela}, {Kawamura},
  {Keih{\"a}nen}, {Keskitalo}, {Kisner}, {Kneissl}, {Knox}, {Kurki-Suonio},
  {Lagache}, {Lamarre}, {Lasenby}, {Laureijs}, {Lawrence}, {Leach}, {Leonardi},
  {Leroy}, {Lilje}, {Linden-V{\o}rnle}, {L{\'o}pez-Caniego}, {Lubin},
  {Mac{\'\i}as-P{\'e}rez}, {MacTavish}, {Maffei}, {Maino}, {Mand olesi},
  {Mann}, {Maris}, {Martin}, {Mart{\'\i}nez-Gonz{\'a}lez}, {Masi}, {Matarrese},
  {Matthai}, {Mazzotta}, {McGehee}, {Meinhold}, {Melchiorri}, {Mendes},
  {Mennella}, {Miville-Desch{\^e}nes}, {Moneti}, {Montier}, {Morgante},
  {Mortlock}, {Munshi}, {Murphy}, {Naselsky}, {Natoli}, {Netterfield},
  {N{\o}rgaard-Nielsen}, {Noviello}, {Novikov}, {Novikov}, {O'Dwyer}, {Onishi},
  {Osborne}, {Pajot}, {Paladini}, {Paradis}, {Pasian}, {Patanchon},
  {Perdereau}, {Perotto}, {Perrotta}, {Piacentini}, {Piat}, {Plaszczynski},
  {Pointecouteau}, {Polenta}, {Ponthieu}, {Poutanen}, {Pr{\'e}zeau}, {Prunet},
  {Puget}, {Reach}, {Reinecke}, {Renault}, {Ricciardi}, {Riller},
  {Ristorcelli}, {Rocha}, {Rosset}, {Rowan-Robinson}, {Rubi{\~n}o-Mart{\'\i}n},
  {Rusholme}, {Sandri}, {Santos}, {Savini}, {Scott}, {Seiffert}, {Shellard},
  {Smoot}, {Starck}, {Stivoli}, {Stolyarov}, {Stompor}, {Sudiwala}, {Sygnet},
  {Tauber}, {Terenzi}, {Toffolatti}, {Tomasi}, {Torre}, {Tristram}, {Tuovinen},
  {Umana}, {Valenziano}, {Vielva}, {Villa}, {Vittorio}, {Wade}, {Wandelt},
  {Wilkinson}, {Yvon}, {Zacchei}, \& {Zonca}}]{Planck2011}
{Planck Collaboration}, {Ade}, P.~A.~R., {Aghanim}, N., {et~al.} 2011, \aap,
  536, A19

\bibitem[{{Reynolds}(1991)}]{Reynolds1991}
{Reynolds}, R.~J. 1991, in The Interstellar Disk-Halo Connection in Galaxies,
  ed. H.~{Bloemen}, Vol. 144, 67

\bibitem[{{Risacher} {et~al.}(2016){Risacher}, {G{\"u}sten}, {Stutzki},
  {H{\"u}bers}, {Bell}, {Buchbender}, {B{\"u}chel}, {Csengeri}, {Graf},
  {Heyminck}, {Higgins}, {Honingh}, {Jacobs}, {Klein}, {Okada}, {Parikka},
  {P{\"u}tz}, {Reyes}, {Ricken}, {Riquelme}, {Simon}, \&
  {Wiesemeyer}}]{Risacher2016}
{Risacher}, C., {G{\"u}sten}, R., {Stutzki}, J., {et~al.} 2016, \aap, 595, A34

\bibitem[{{Rolleston} {et~al.}(2000){Rolleston}, {Smartt}, {Dufton}, \&
  {Ryans}}]{Rolleston2000}
{Rolleston}, W.~R.~J., {Smartt}, S.~J., {Dufton}, P.~L., \& {Ryans}, R.~S.~I.
  2000, \aap, 363, 537

\bibitem[{{Rubin}(1989)}]{Rubin1989}
{Rubin}, R.~H. 1989, \apjs, 69, 897

\bibitem[{{Samson} \& {Angel}(1990)}]{Samson1990}
{Samson}, J.~A.~R. \& {Angel}, G.~C. 1990, \pra, 42, 1307

\bibitem[{{Sch{\"o}ier} {et~al.}(2005){Sch{\"o}ier}, {van der Tak}, {van
  Dishoeck}, \& {Black}}]{Schoier2005}
{Sch{\"o}ier}, F.~L., {van der Tak}, F.~F.~S., {van Dishoeck}, E.~F., \&
  {Black}, J.~H. 2005, \aap, 432, 369

\bibitem[{{Sternberg} {et~al.}(2003){Sternberg}, {Hoffmann}, \&
  {Pauldrach}}]{Sternberg2003}
{Sternberg}, A., {Hoffmann}, T.~L., \& {Pauldrach}, A.~W.~A. 2003, \apj, 599,
  1333

\bibitem[{{Tayal}(2008)}]{Tayal2008}
{Tayal}, S.~S. 2008, \aap, 486, 629

\bibitem[{{Tayal}(2011)}]{Tayal2011}
{Tayal}, S.~S. 2011, \apjs, 195, 12

\bibitem[{{Velusamy} {et~al.}(2015){Velusamy}, {Langer}, {Goldsmith}, \&
  {Pineda}}]{Velusamy2015}
{Velusamy}, T., {Langer}, W.~D., {Goldsmith}, P.~F., \& {Pineda}, J.~L. 2015,
  \aap, 578, A135

\bibitem[{{Velusamy} {et~al.}(2012){Velusamy}, {Langer}, {Pineda}, \&
  {Goldsmith}}]{Velusamy2012}
{Velusamy}, T., {Langer}, W.~D., {Pineda}, J.~L., \& {Goldsmith}, P.~F. 2012,
  \aap, 541, L10

\bibitem[{{Voronov}(1997)}]{Voronov1997}
{Voronov}, G.~S. 1997, Atomic Data and Nuclear Data Tables, 65, 1

\bibitem[{{Young} {et~al.}(2012){Young}, {Becklin}, {Marcum}, {Roellig}, {De
  Buizer}, {Herter}, {G{\"u}sten}, {Dunham}, {Temi}, {Andersson}, {Backman},
  {Burgdorf}, {Caroff}, {Casey}, {Davidson}, {Erickson}, {Gehrz}, {Harper},
  {Harvey}, {Helton}, {Horner}, {Howard}, {Klein}, {Krabbe}, {McLean}, {Meyer},
  {Miles}, {Morris}, {Reach}, {Rho}, {Richter}, {Roeser}, {Sandell}, {Sankrit},
  {Savage}, {Smith}, {Shuping}, {Vacca}, {Vaillancourt}, {Wolf}, \&
  {Zinnecker}}]{Young2012}
{Young}, E.~T., {Becklin}, E.~E., {Marcum}, P.~M., {et~al.} 2012, \apjl, 749,
  L17

\end{thebibliography}


\end{document}